\documentclass[pra,twocolumn,superscriptaddress,10pt]{revtex4-1}
\usepackage{graphicx}
\usepackage{dcolumn}
\usepackage{color}
\usepackage{times}
\usepackage{bm}
\usepackage{amssymb}
\usepackage{amsmath}
\usepackage{epsfig}
\usepackage{epstopdf}
\usepackage{dsfont}
\usepackage{float}
\usepackage{subfigure}
\RequirePackage[colorlinks,citecolor=blue,urlcolor=blue,linkcolor=blue]{hyperref}
\begin{document}
	\title{Characterization of  the measurement uncertainty dynamics in an open system}
	\author{Xi-Hao Fang}
	\affiliation{\rm School of Physics and Astronomy, Sun Yat-Sen University, Guangzhou 510275,People's Republic of China}
\author{Fei Ming}
	\affiliation{\rm School of Physics \& Material Science, Anhui University, Hefei
230601,  People's Republic of China}
\author{Dong Wang}\email{dang@ahu.edu.cn}
	\affiliation{\rm School of Physics \& Material Science, Anhui University, Hefei
230601,  People's Republic of China}
	\affiliation{\rm CAS  Key  Laboratory  of
Quantum  Information,  \\ University  of  Science  and
Technology of China, Hefei 230026,  People's Republic of China}

	\begin{abstract}
	Uncertainty principle plays a crucial role in quantum mechanics, because it captures the essence of the inevitable randomness associated with the outcomes of two incompatible quantum measurements. Information entropy can perfectly describe this type of randomness in information theory, because entropy can measure the degree of chaos in a given quantum system. However, the quantum-assisted uncertainty of entropy  eventually inflate inevitably as the quantum correlations of the system are progressively corrupted by noise from the surrounding environment. In this paper, we investigate the dynamical features of the von Neumann entropic uncertainty in the presence of quantum memory, exploring the time evolution of entropic uncertainty suffer noise from the surrounding. 	 Noteworthily, how the environmental noises affect the uncertainty of entropy is revealed, and specifically we verify how two types of noise environments, AD channel and BPF channel, influence the uncertainty. Meanwhile, we put forward some effective operation	strategies to reduce the magnitude of the measurement uncertainty under the open systems. Furthermore, we explore the applications of the uncertainty relation investigated on entanglement witness and quantum channel capacity. Therefore  , in open  quantum correlation systems, our investigations could provide an insight into quantum measurement estimation
	\end{abstract}

\maketitle

	\section{  Introduction}
	The Heisenberg uncertainty relation is one of the fundamental concepts in quantum theory, first proposed by German physicist Heisenberg in 1927 \cite{1}. In quantum mechanics, wave function $\psi$ is used to describe the state of particles, and the square of the wave function   $\vert\psi(x)\vert^2$ describes the probability that the particlebe found somewhere. After the probability of wave of function in quantum mechanics is proposed, quantum mechanics partially preserves the concept of waves and particles, and the uncertainty principle sets limit on predicting the measurement outcomes of two incompatible observables. Kennard \cite{2} formalized Heisenberg original ideas in an uncertainty relation by a pair of observables:
	\begin{equation}
	\triangle p_x\cdot\triangle x\geq\frac{\hbar}{2}
	\label{E1}
	\end{equation}
	 $p_x$ and $x$ are momentum and position of one particle, where   uncertainties in the measurements are quantified in terms of the standard deviations.
	\begin{equation}
	 \triangle x=\sqrt{\left\langle x^{2}\right\rangle-\left\langle x\right\rangle^{2}} \qquad     \triangle p_x=\sqrt{\left\langle p_x^{2}\right\rangle-\left\langle p_x\right\rangle^{2}},
	 	\label{E2}
	 \end{equation}
	
	 Soon after, Robertson \cite{3} and Schrodinger \cite{4} generalized it to arbitrary pairs of noncommuting observables $P$ and $Q$ and it is well known in the following form:
	 	\begin{equation}
	 \triangle Q\cdot\triangle P\geq\frac{1}{2}\vert\left\langle Q,P\right\rangle\vert,
	 	\label{E3}
	 \end{equation}
	 uncertainties still quantified in terms of the standard deviations
	 	\begin{equation}
	\triangle x=\sqrt{\left\langle Q^{2}\right\rangle-\left\langle Q\right\rangle^{2}} \qquad     \triangle P=\sqrt{\left\langle P^{2}\right\rangle-\left\langle P\right\rangle^{2}}.
		\label{E4}
	\end{equation}

	Afterward ,  for characterizing the "spread" in the outcomes of two incompatible measurements was  advocate dy	Deutsch \cite{5} . In recent years, uncertainty relationship develops rapidly on entropy of information. Here, suppose we use denote $p_{i}(x)$ the probability of the outcome, the Shannon entropy can be expressed as $ H(x)=-\sum_	i p_{i}\log_{2}p_{i}(x) $. It indicates the uncertainty of a system, large entropy represents large uncertainty. When there are two random variables, mutual information, joint entropy, conditional entropy can be used to describe the relationship between the information they contain. Let the joint probability distribution of the observables $x$ and $y$ is
	
	\begin{equation}
	 H(X,Y)=-\sum_{x,y} p(x,y)\log_{2}p(x,y) .
	 	\label{E5}
	\end{equation}

	Conditional Shannon entropy is a generalization of conditional probability, the uncertainty in system $ (x,y) $ Conditional Shannon entropy is 
	
	\begin{equation}
	H(x\vert y)=H(x,y)-H(y) . \\
		\label{E6}
	\end{equation}
	
	In the bipartite state,  von Neumann entropy replac Shannon entropy as
	
	\begin{equation}
	S(\rho)=-Tr(\rho \log_2\rho)=-\sum_x\lambda_x\log_2\lambda_x ,
		\label{E7}
	\end{equation}
	it also represents the uncertainty of the quantum system, $ \lambda_x$ is the  eigenvalue of observables. At first, Bialynicki-Birula \cite{6} proved the entropic uncertainty relation(EUR) of observable position and momentum entropy in 1975. Another conjecture put forth by Kraus  \cite{7}  was proved by Maassen and Uffink  \cite{8}, leading to the following EUR fomular:
	
	\begin{equation}
	H(P)+H(Q) \geq \log_2\frac{1}{c} :=q_{MU} ,
		\label{E8}
    \end{equation}	
    	here $ c=\rm max_{i,j}c_{i,j} $ and $ c_{i,j}=\vert\left\langle {x_i\vert z_i}\right\rangle\vert^{2}$.

It is remarkably noting that the uncertainty principl can be described based on an interesting game model between two participants  Alice and Bob. In the game, Bob prepares a particle $A$ being entangled with his quantum memory $B$ and sends it to Alice, and then Alice carries out two measurements $Q$ and $P$ on the particle $A$ and announces her choice to Bob. Bob's task is to minimize his uncertainty about measurement outcome of Alice. Mathematically, the relation can be expressed as \cite{9,10,11,12,13,14,15,16,17}
	
		\begin{equation}
	 S(P\vert B)+S(Q\vert B) \geq \log_2\frac{1}{c}+S(A\vert B)=U_B ,
	 \label{E9}
	 \end{equation}
	  where $S(K\vert B)=S(\rho_{KB})-S(\rho_B)$ is the post-measurement state $\rho_{AB}$ after quantum system $ A $  is measured in K-basis, the memory \cite{18}
	
	\begin{equation}
	\rho_{XB}=\sum_x(\vert\varphi_x\rangle\langle\varphi_x\vert \otimes I)\rho_{AB}(\vert\varphi_x\rangle\langle\varphi_x\vert\otimes I),	
		\label{E10}
	 \end{equation}
    respectively resulting after the POVMs  $ P$ and $Q$ are performed on the system $A$.
	
	 In this scheme, we use two Pauli observables  and  as the measurement. Hence, it is easy to obtain $c=\frac{1}{2}$  in this case.
	
	 Years later, several authors improved and tightened this bound , Pati \cite{19} proved that the uncertainties are lower bounded by an additional term compared to Eq. (\ref{E9}) as
	
	 \begin{align}
 S(X\vert B)+S(Z\vert B) \geq &\log_2\frac{1}{c}+S(A\vert B)\nonumber\\
 &+\max\{0,D_A(\rho^{AB})-J_A(\rho^{AB})\}
\label{E11}
	\end{align}

	 $ J_A(\rho^{AB}) $ is the classical correlation, it can defined as
	
	 	\begin{equation}
	 	J  _A(\rho^{AB})\equiv \max_{\Pi_k}J(B\vert{\Pi_k}),
	 		\label{E12}
	 	\end{equation}
	  where the optimization	over all positive operator-valued measures acting on measured particle $A$. The difference between the total and the classical correlations can be expressed by quantum discord (QD). \cite{20,21,22,23}
	
	 	\begin{equation}
	 D_A(\rho^{AB})=I(\rho^{AB})-J_(\rho^{AB}).
	 	\label{E13}
	 \end{equation}
	where $ I(a:b) $ called mutual information, representing the information correlation  between $b$ and $a$
	
	  	\begin{equation}
	  I(A:B)=H(a)+H(b)-H(a,b),
	  	\label{E14}
	  \end{equation}
	   total correlation is
	
	   	\begin{equation}
	   I(\rho_{AB})=S(\rho_{A})+S(\rho_B)-S(\rho_{AB}),
	   	\label{E15}
	   \end{equation}
	 
	  The lower bound in Eq. (\ref{E11}) tightens the bound in Eq. (\ref{E9}) if the discord $ D_A(\rho^{{AB}}) $  is larger than the classical correlation $ J_A(\rho^{{AB}}) $.
	
	  Very recently, F. Adabi and S.Salimi \cite{24} further improved the uncertainty relationship based on the mutual information of the two-particle state . Consider a bipartite state $ \rho_{AB} $ shared between Alice and Bob. Alice performs the $ X $ or $ Z $ measurement and announce her choice to Bob. Bob's uncertainty about both $ X $ and $ Z$   measurement outcomes is
	
	\begin{equation}
    S(X\vert B)+S(Z\vert B) \geq q_{UM}+S(A\vert B)+\max\{0,\delta\} ,
    	\label{E17}
    \end{equation}
	  where
	  	\begin{equation}
 \delta=I(a:b)-[I(X:b)+I(Z:b)].
 	\label{E18}
	  \end{equation}
	
	  When Alice measures observable $P$ on her particle, she will obtain the $i$ th outcome with probabilityand Bob's particle will be left in the corresponding state $ \rho_{B}^k=\frac{Tr_{AB}(\Pi^A_{i}\rho\Pi^A_i)}{p_i} $.
	
	  Then the Holevo quantity \cite{25} is
	  \begin{equation}
	  I(P:b)=S(\rho^B)-\sum_ip_iS(\rho^B_i),
	  	\label{E19}
	  \end{equation}
	  it is equal to the upper bound of Bob's accessible information about Alice's measurement results. Thereby, It can be seen that if the sum of the information sent by Alice to Bob through her measurements is less than the mutual information between $A$ and $B$, the above EUR represents an improvement to Berta's uncertainty relation by raising the lower bound limit by the amount of . It is worth noting that  if observables  and  are complementary and subsystem $A$ is a maximally mixed state ,the inequality Eq. (\ref{E9}) becomes an equality.  Thus, for  the class of states   maximally mixed subsystem $A$ (including Werner states, Bell diagonal states, and isotropic states) ,the lower bound is perfectly tight.

	  This article's layout  provided as follow. In Section \ref{S2}, under two different channel, we investigate  the dynamical traits  of the entropic uncertainty and compare several different lower bounds Eq. (\ref{E9},\ref{E11},\ref{E17}). The results show that the dynamic of uncertainty will exhibit non-monotonic behaviors over time. In addition, we find that the lower bound proposed by Adabi's bound is tighter than the  bound proposed  by Berta and Pati. In Section \ref{S3}, by deriving the optimal combination between local amplitude damping and bit-phase flip channel, we offer two feasible  and effective strategies to steer the measurement uncertainty as a quantum weak measurement and filtering operation. More over, in Section \ref{S4},  we give two functional applications of the test uncertainty relationship to entanglement witness and quantum channel capacity. At last, we summarize our findings in Section \ref{S5}.
	
	\section{  Entropy Quantum-memory-assisted entropic uncertainty under environment noise}
	\label{S2}
	
	In order to further explore the uncertainty in different environments noise, let us return to the uncertain game. Bob sends $A$ particle which is affected by environmental noise to Alice initially,   and  particles $B$ remains as it is because it's not in the sending process. Kraus \cite{7} used operators $ E_i $ to describe the process that the particle $ A $ to be measured  is exposed to the environment characterized by evolution, and usually $ B $ as quantum memory so it is very robust and is not affected by any decoherence. Therefore, we can give  the evolution state of the system as
		\begin{equation}
	\rho_{AB}(t)=T(\rho_{AB})=\sum_i(E_i \otimes I^B)\rho_{AB}(E_i\otimes I^B)^\dagger.
		\label{E20}	
	\end{equation}
	
  In the light of quantum-memory-assisted game mod, Bob hold a pair of particles $A$ and $B$, and also $A$ and $B$ are entangled initially in  the form of Bell-diagonal state
	
		\begin{equation}
	\rho_{AB}=(I^A\otimes I^B +\sum_{j=1}^{3}C_{\sigma j}\sigma^A_j\otimes \sigma^B_j),
	\label{E21}
	\end{equation}
	where $\sigma$ are the standard Pauli matrices, $C_{\sigma j}=Tr_{AB}(\rho_{AB}\sigma^A_j\otimes\sigma^B_j)$ and the coefficients of  initial systematic state with  $0\leq\vert C_{\sigma j}\vert\leq1$  , $A$ is maximally entangled with $B$.
	\subsection{ SPMC condition  }
	The right-hand side of Quantum-memory-assisted entropic uncertainty is
	
		\begin{equation}
U_b=\log_2\frac{1}{c}+S(A|B).
\label{E22}
	\end{equation}
	
	If we measureing  the  uncertainty of $\sigma_j$ and $\sigma_k$, then when the initial quantum state  satisfy SPMC condition \cite{26}

	\begin{equation}
      C_{\sigma i}=-C_{\sigma j}C_{\sigma k}(i\ne j \ne k),
      \label{E23}
	\end{equation}	
	
       That is	$U\equiv U_b, $ the entropic uncertainty relation can always reach the bound. 
	\subsection{ Amplitude damping channel }

     Amplitude damping channel is a description of the dissipation of energy in a quantum system , the Kraus \cite{12} operators can be given in sum-operator representation as
	 \begin{equation}
	E^{AD}_1= \left(
	\begin{array}{cc}
	1  & 0  \\
	0  & \sqrt{1-d} \\
	\end{array}
	\right)	
	\qquad
	E^{AD}_2= \left(
	\begin{array}{cc}
	0 & \sqrt{d} \\
	0 &    0
	\end{array}
	\right).
	\label{E24}
	\end{equation}

	Where $d=1-e^{-\lambda t}$ represents the parameter of the energy dissipation rate, and $\lambda$ denotes the energy relaxation rate. time value $ t $ range is $[0,1]$ , under this channel, inital state $ \vert 0 \rangle $ amplitude will decrease with time. Amplitude damping channel is a spontaneous emission model with zero  ambient temperature, describe the energy consumption.

	After go through the  above environmental noise mentioned above, the matrix elements  of the bipartite state $	\rho_{AB}(t)$ are taken as	$\rho_{11}=\frac{1}{4}(1 + C_3 + d - C_3 d),	\rho_{22}=\frac{1}{4}(1 + C_3 (-1 + d) + d),   \rho_{33}=\frac{1}{4}(-1 + C_3) (-1 + d),	\rho_{44}=-\frac{1}{4} (1 + C_3) (-1 + d),	\rho_{14}=\rho_{41}=\frac{1}{4}(C_1 - C_2) \sqrt{1 - d},	\rho_{23}=\rho_{23}=\frac{1}{4}(C_1 + C_2) \sqrt{1 - d}. $
		
			 Four eigenvalues of density matrix are
			\begin{align}
		\lambda^{AD}_1&=\frac{1}{4}(1 - C_3 + C_3 d \nonumber\\
			&-\sqrt{C_1^2 + 2 C_1 C_2 + C_2^2 - C_1^2 d - 2 C_1 C_2 d - C_2^2 d + d^2}),\nonumber\\
		\lambda^{AD}_2&=\frac{1}{4}(1 - C_3 + C_3 d \nonumber\\
			&+\sqrt{C_1^2 + 2 C_1 C_2 + C_2^2 - C_1^2 d - 2 C_1 C_2 d - C_2^2 d + d^2}),\nonumber\\
		\lambda^{AD}_3&=\frac{1}{4}(1 + C_3 - C_3 d  \nonumber\\
			&-\sqrt{C_1^2 - 2 C_1 C_2 + C_2^2 - C_1^2 d + 2 C_1 C_2 d - C_2^2 d + d^2}),\nonumber\\
		\lambda^{AD}_4&=\frac{1}{4}(1 + C_3 - C_3 d  \nonumber\\
			&+\sqrt{C_1^2 - 2 C_1 C_2 + C_2^2 - C_1^2 d + 2 C_1 C_2 d - C_2^2 d + d^2}),
			\label{E25}
			\end{align}

	To illustrate the feature of uncertainty, we may resort to $ \sigma_x $ and $ \sigma_z $ as the measurement's incompatibility, which is always used for describing the spin $ -\frac{1}{2} $ observables, with the eigenstates of $\vert\sigma_x^{\pm}\rangle=\frac{(1,\pm1)^T}{\sqrt{2}},\vert\sigma_z^{\pm}\rangle={(1,0)^T}.$
	
	Therefore, the conditional von Neumann entropies representing the uncertainty of the measurement outcomes are respectively given by $S(\sigma_x\vert B)=S(\rho_{\sigma_x})-S(\rho_B)$ and $S(\sigma_z\vert B)=S(\rho_{\sigma_z})-S(\rho_B).$
 Thus, the left-hand side of Eq. (\ref{E9}) can be given by
 	\begin{align}
  U= &-\frac{1}{4}\{4\mu_1 \rm arctanh (\mu_1)+2\mu_2\rm arctanh(\mu_2) \nonumber\\
 &-2\mu_3\rm arctanh(\mu_3)+2\log_2(1+\mu_1)+2\log_2(1+\mu_2)\nonumber\\
 &+\log_2(1-\mu_2)+\log_2(1+\mu_2)+\log_2(1+\mu_3)\nonumber\\
 &+\log_2(1-\mu_3)\}
 \label{E26}
 \end{align}

 	where $\mu_1=\sqrt{-(C_1 - C_2) (C_1 + C_2) (-1 + d)}, \mu_2=C_3 + d - C_3 d,\mu_3=C_3-C_3d-d.$

	On the other hand, if the reduced density matrix of the Bell diagonal state maintains  maximally mixed ,the complementarity $c$ of the observables $x$ and $z$ is exactly equal to $\frac{1}{2}$. Thus, the right-hand side of Eq. (\ref{E9}) takes the form
		
	\begin{equation}
U_B=\log_{2}\frac{1}{c}+S(A\vert B)=S(\rho_{AB})-S(\rho)+1,
\label{E25}
	\end{equation}
	
\begin{figure*}
\centering
{\includegraphics[width=7cm]{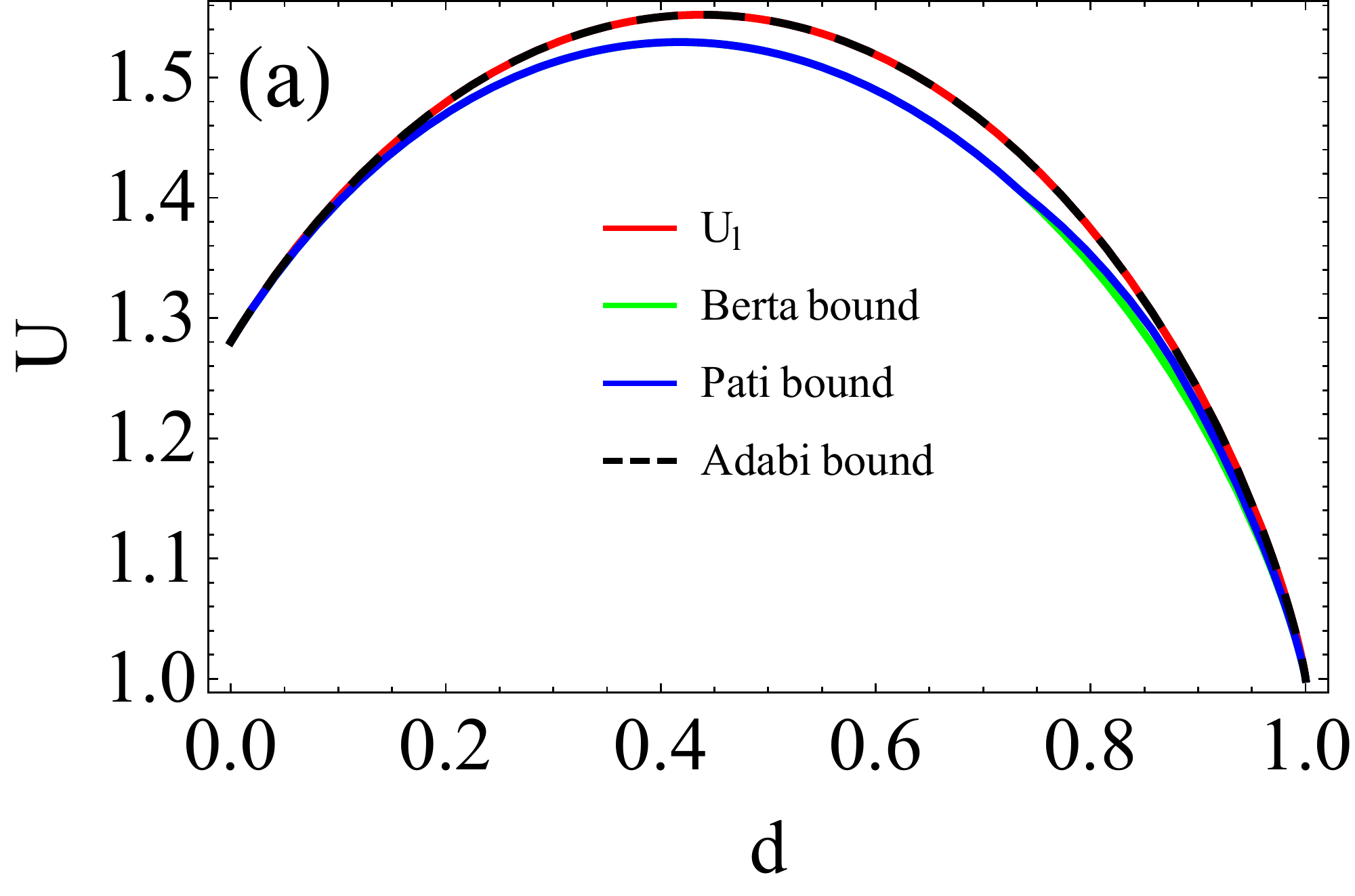}} \ \ \ \
{\includegraphics[width=7cm]{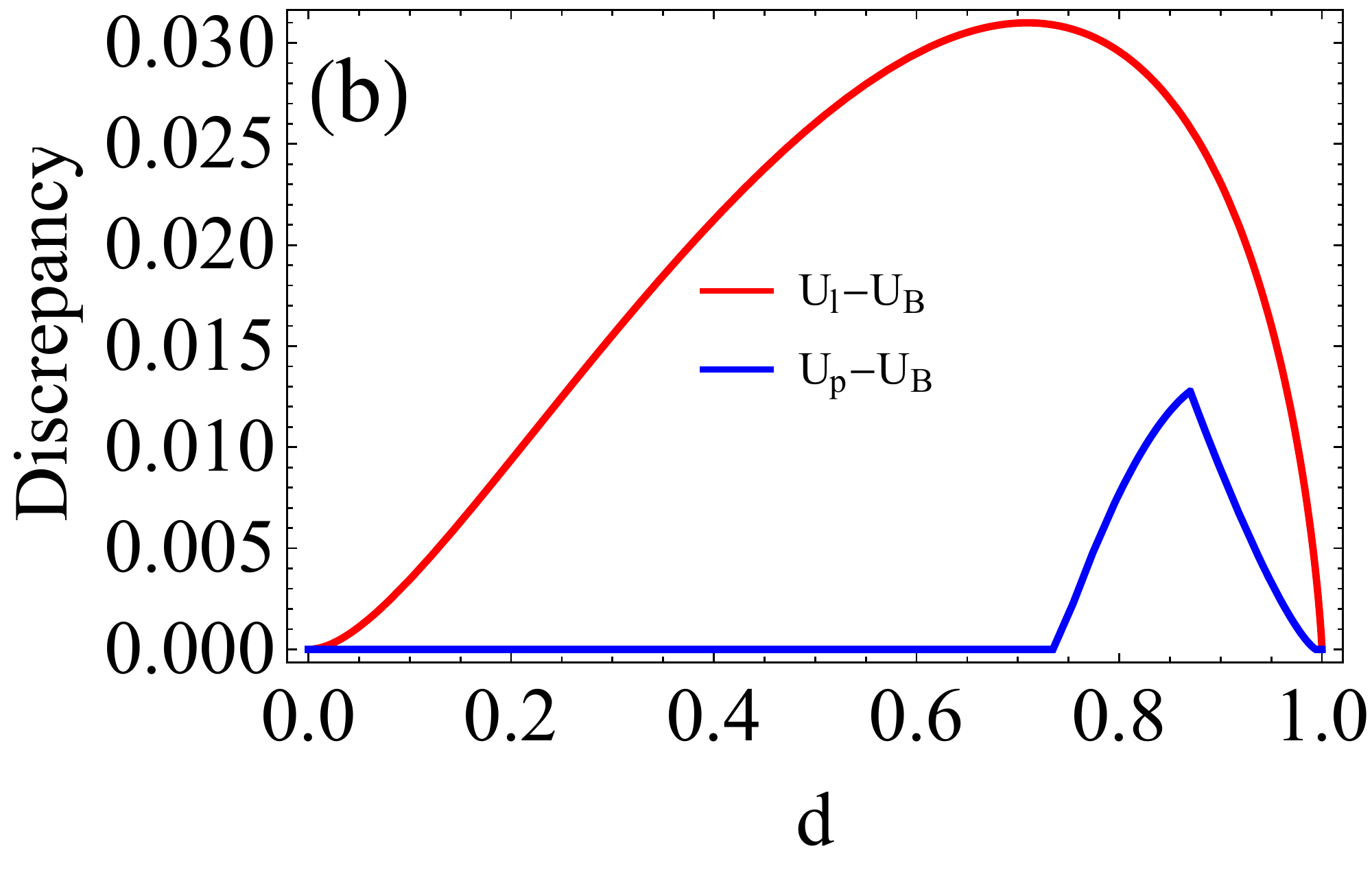}}
{\includegraphics[width=7cm]{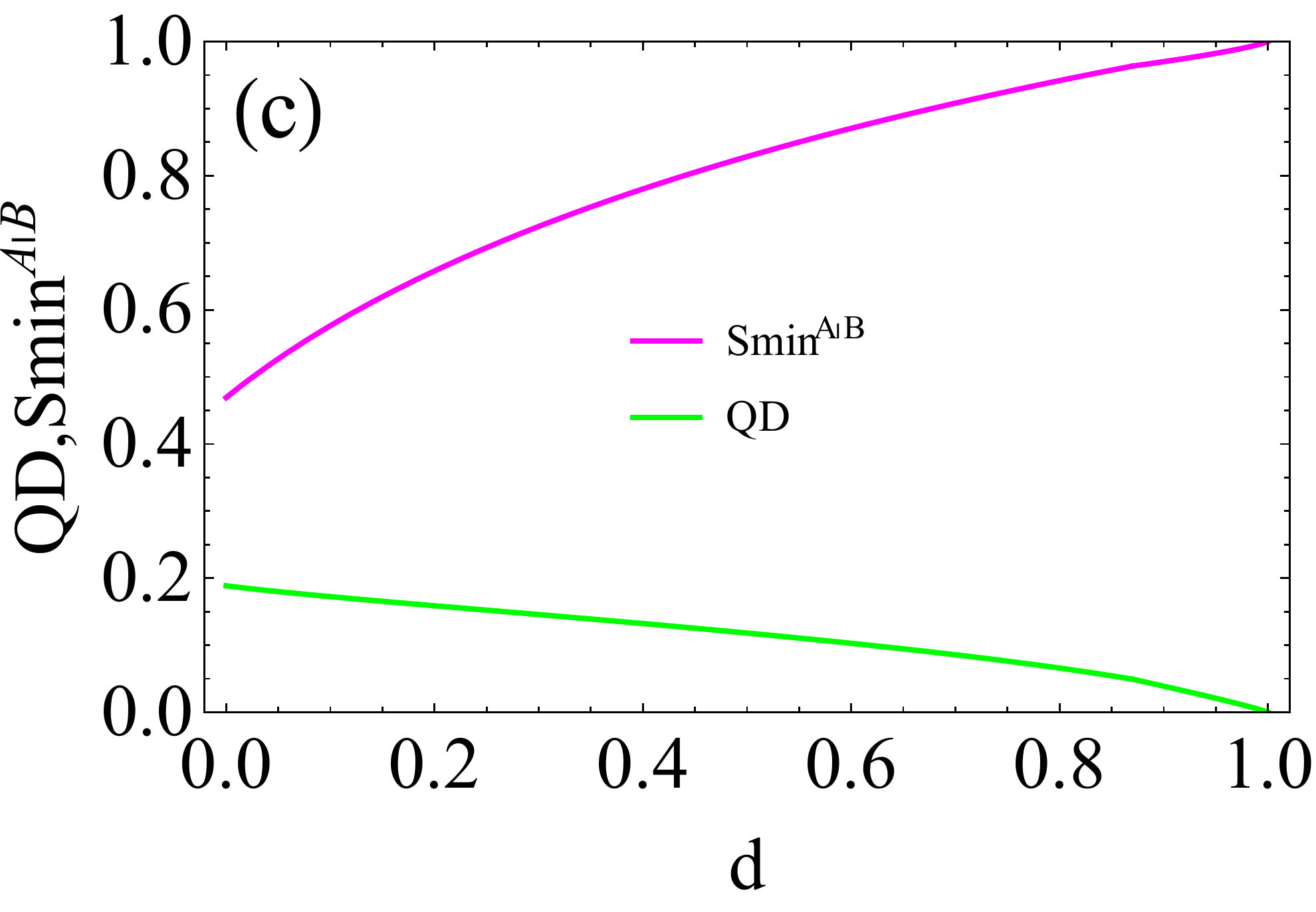}} \ \ \ \
{\includegraphics[width=7cm]{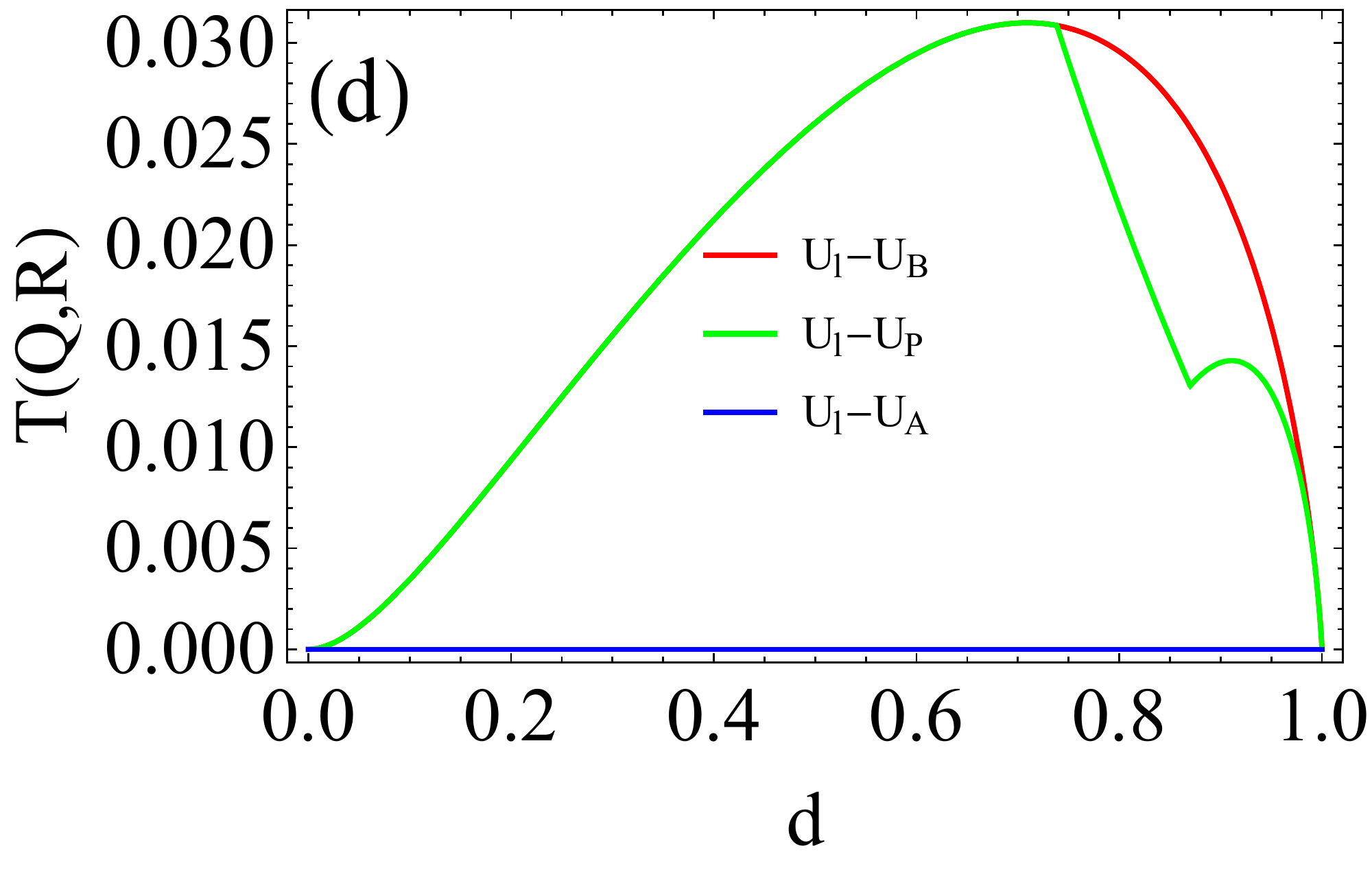}}
\caption{Dynamics of entropic uncertainty under AD channel, several kinds of lower bound and quantum correlation of the system  with respect to the energy dissipation rate $d=1-e^{-\lambda t}$ when the initial state is prepared with $ C_1=-0.5,C_2=0.4,C_3=0.8 .$ Graph (a): $U_l$ denotes the entropic uncertainty expressed in Eq. (\ref{E9}) , the yellow line denotes the bound proposed by Berta et al., the blue line denotes the Pati et al.'s bound, the black dashed line represents the bound by Adabi et al. The  graph (b) and(d): $U_A$ denotes the Adabi et al.'s lower bound; $U_B$ denotes the Berta et al.'s bound, $U_P$ denotes the Pati et al.'s bound. Graph (c):QD denotes quantum discord expressed in  Eq. (\ref{E9}) , $S^{A\vert B}_{\rm min}$ denotes the $A$'s minimal condition entropy in Eq. (\ref{E26}). }

	\label{fig:1}
\end{figure*}

	In order to show traightly  the evolution of entropy uncertain relation, we show the diagrams of the uncertainty and the existed different bounds as functions of the time-parameter $d=1-e^{-\lambda t}$ with $C_1=-0.5,C_2=0.4,C_3=0.8$, in Figure \ref{fig:1}. From the figure, we can find that the entropic uncertainty will resonantly reduce with the growing time $t$.
	
	Obviously, with the  $t$ increase, the measurement uncertainty and  the referred bounds will reveal non-monotonic feature and  gradually close.In addition, it is interesting to find that the  bound of  Adabi et al is perfectly coincident with the left-hand side of the uncertainty expressed in Eq. (\ref{E9}). And one can see that Pati et al's bound is tighter than that proposed  by Berta et al, and Adabi et al's bound is tighter than Pati et al's one, as shown in Figure \ref{fig:1}b. This states that Adabi et al's bound is optimal compared with the others, which is, $U_B\leq U_P\leq U_A $ is hold.
%

	In addition, QD has been shown in Figure \ref{fig:1}c , it can quantify the dynamic of the system's quantum correlation under AD channel.  Mathematically, in the current  bipartite system with an arbitrary X-structure density matrix, the QD of the system can be accurately derived as
	\begin{equation}
	D(\rho_{AB})=S_{\rm bin}(\rho_{22}+\rho_{44}+\sum_{i=1}^{4}\lambda_i\log_2\lambda_i+\min\{P_1,P_2\}),
	\label{E27}
	\end{equation}
	with
		\begin{equation}
	P_1=S_{\rm bin}(\Gamma),P_2=-\sum_{n}^{i=1}\rho_{ii}\log_2\rho_{ii}-S_{\rm bin}(\rho_{11}+\rho_{33}).
	\label{E28}
	\end{equation}
	where $\Gamma=\frac{1}{2}\{1+\sqrt{[1-2(\rho_{33}+\rho_{44})]^2+4(\vert\rho_{14}\vert+\vert\rho_{23}\vert)^2}\} $, $\rho_{ij}$ is the element of the corresponding density matrix and $\lambda_i$ denotes the eigenstates of $\rho_{AB} $. 
	
	We obviously find  that this is counter-intuitive in the current situation, that the bigger quantum correlation  not induce the smaller uncertainty, as shown in Figure \ref{fig:1}c. In the following, we will	render a physical explanation for this phenomenon. To	interpret this abnormal  behavior, we can combine	Eq. (\ref{E9}) and Eq. (\ref{E13}), and it  can derive that the QD is	equal to
		
	\begin{equation}
	U_B=1+S^{A\vert B}_{\min}-D(\rho_{AB})
	\label{E26}
	\end{equation}
		where $ S^{A\vert B}_{\min}=\min_{B_i}\sum_iq_iS(\rho^A_{i})$ is  $A$'s minimal condition entropy.  This directly indicates that the uncertainty's bound of interest is not only determined by $ D(\rho_{AB}) $, and also by $ S^{A\vert B}_{\min}.$

	 We define the discrepancy between the entropic uncertainty and the lower bound as tightness ,itcan further explore the different bounds presented, let us turn to focus on the differences between the entropic uncertainty and the existed bounds,tightness define as

		\begin{equation}
 T(Q,R)=U_l-U_r,
\label{E29}
	\end{equation}
	where $U_l$  denotes the left-hand-side of the uncertainty relation in Eq. (\ref{E9}), and $U_r$  denotes the right-handside of the relation in Eq. (\ref{E9}), (\ref{E11}) and (\ref{E17}). Therefore, Figure \ref{fig:1}d shows the tightness $T(Q,R)$ for these uncertainty relations with the increasing time $t$. accroding the figure, we obtain the magnitude of the Berta et al's tightness is larger than Pati et al's tightness and is larger than  Adabi et al's tightness , compared with the other bounds ,Adabi et al's bound is tightest.  Moreover, it  directly shows that the tightness of Adabi et al  is always maintained at zero,in other words, all the relations'tightness  mentioned above, except which proposed by Adabi, are used to never fully synchronize with the measurement uncertainty.
	
	\subsection{ Bit-phase flip channel }
	Bit flip, phase flip and bit-phase flip (BPF) also called $\sigma_x$ error, $\sigma_y  $ error and  $\sigma_z  $ error, generally used to describe the effects of quantum systems on flipped noise: loss of information may occur during the process but there is no dissipation of energy. After the particles $A$ pass through the BPF channel, the density matrix still maintains maximally entangled state. the Kraus \cite{7} operators is
	
	  \begin{equation}
	E^{BPF}_1= \sqrt{p}\left(
	\begin{array}{cc}
	1  & 0  \\
	0  & 1 \\
	\end{array}
	\right)	
	\qquad
E^{BPF}_2= \sqrt{1-p}\left(
	\begin{array}{cc}
	0 & 1 \\
	1 &    0
	\end{array}
	\right).
	\label{E30}
	\end{equation}
	
	  Single-particle bits not change with a probability of  $ \sqrt{1-p}$ , change with a probability of $p, $   where $p$  is the time parameter, value range also is $[0,1]$ .
	
	The state of the composite system consisting of AB would take a matrix form of $\rho_{11}=\frac{1}{4}(1 + C_3 (-1 + 2 p)), 	\rho_{22}=\frac{1}{4}(1 + C_3 - 2 C_3 p),	  \rho_{33}=\frac{1}{4}(1 + C_3 - 2 C_3 p),	\rho_{44}=-\frac{1}{4}(1 + C_3 (-1 + 2 p)),	       \rho_{14}=\rho_{41}=\frac{1}{4} (-C_2 + C_1 (-1 + 2 p	)),	\rho_{23}=\rho_{23}=\frac{1}{4}(C_2 + C_1 (-1 + 2 p)).$
	
%
	
\begin{figure*}
\centering
{\includegraphics[width=7cm]{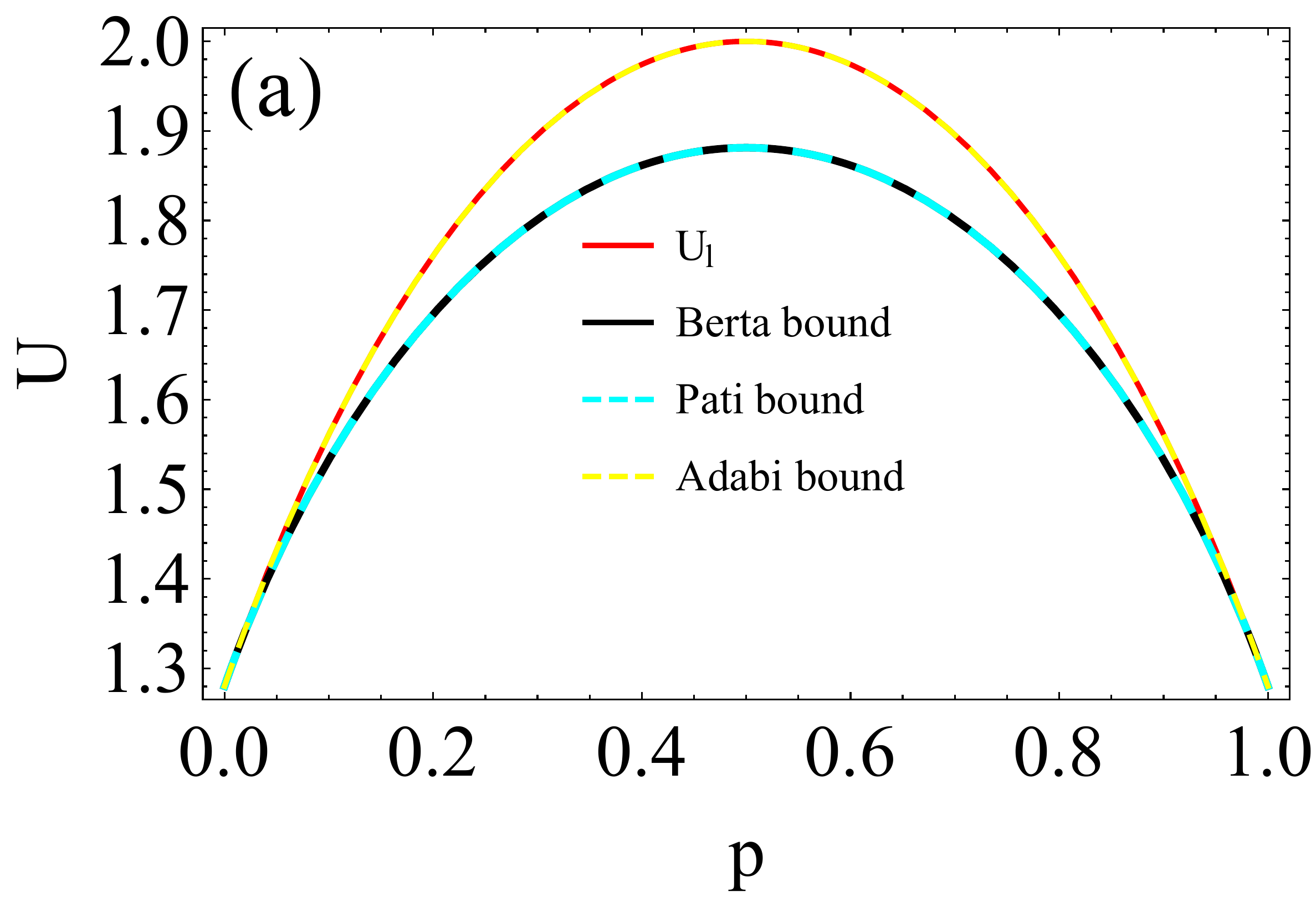}} \ \ \ \
{\includegraphics[width=7cm]{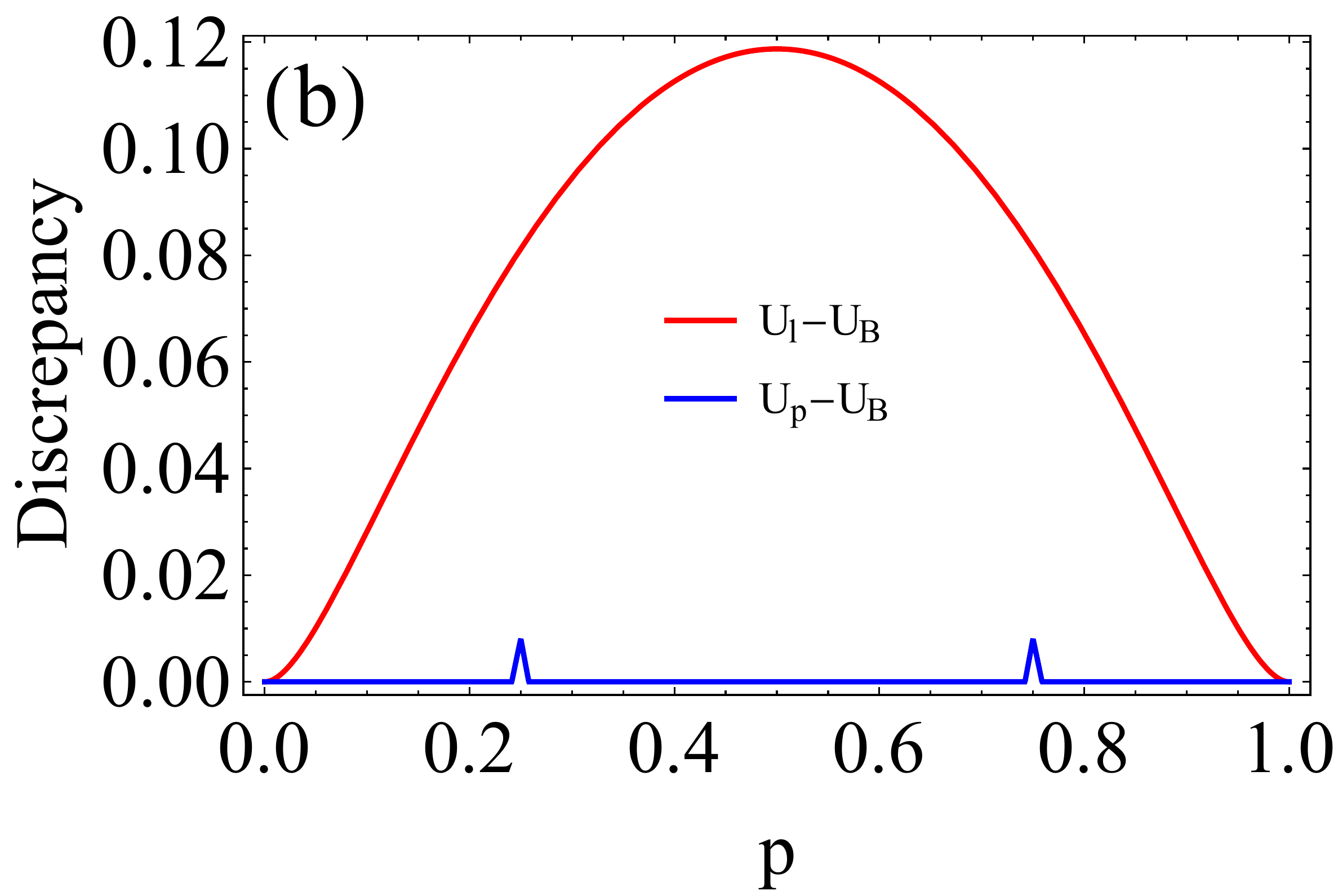}}
{\includegraphics[width=7cm]{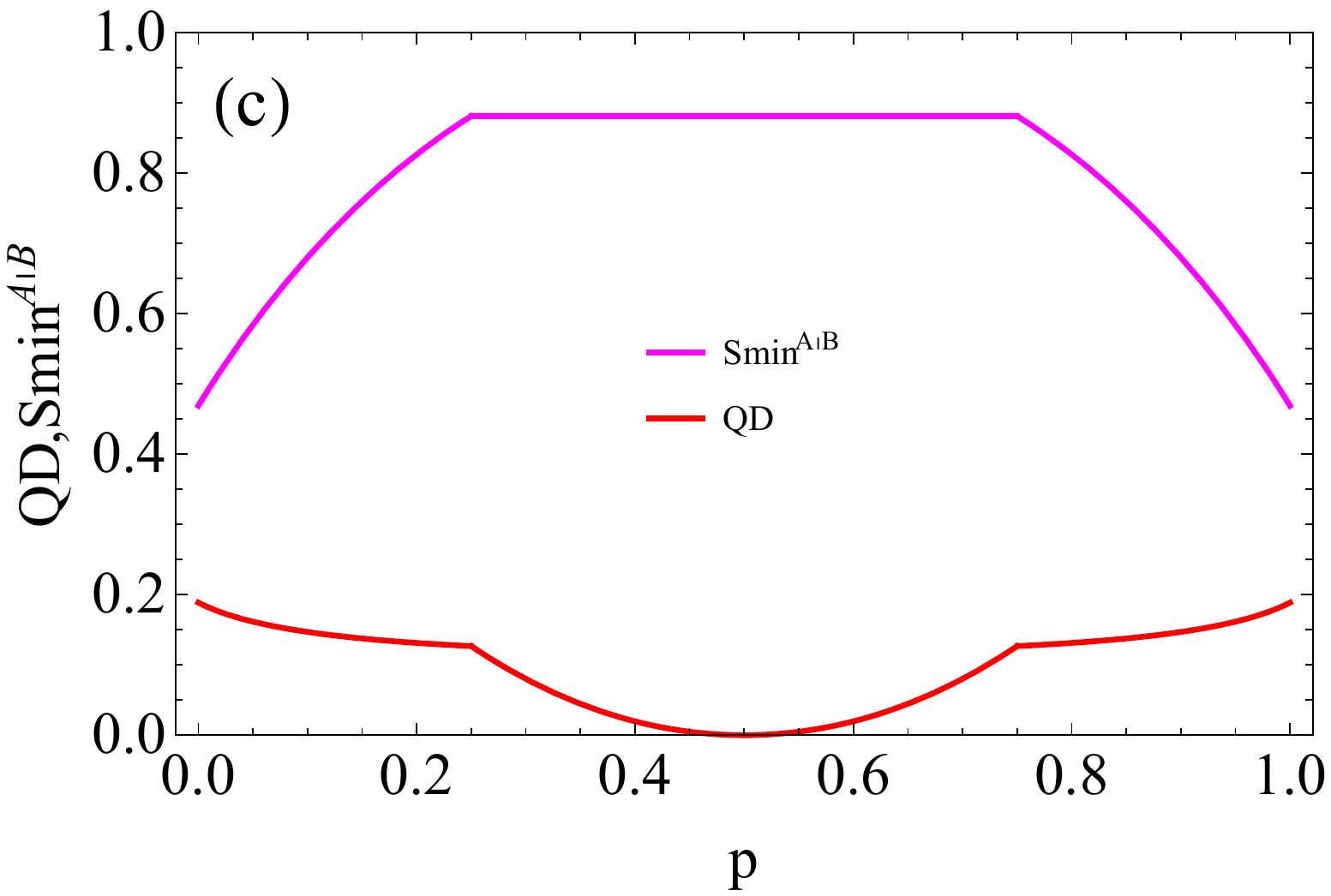}} \ \ \ \
{\includegraphics[width=7cm]{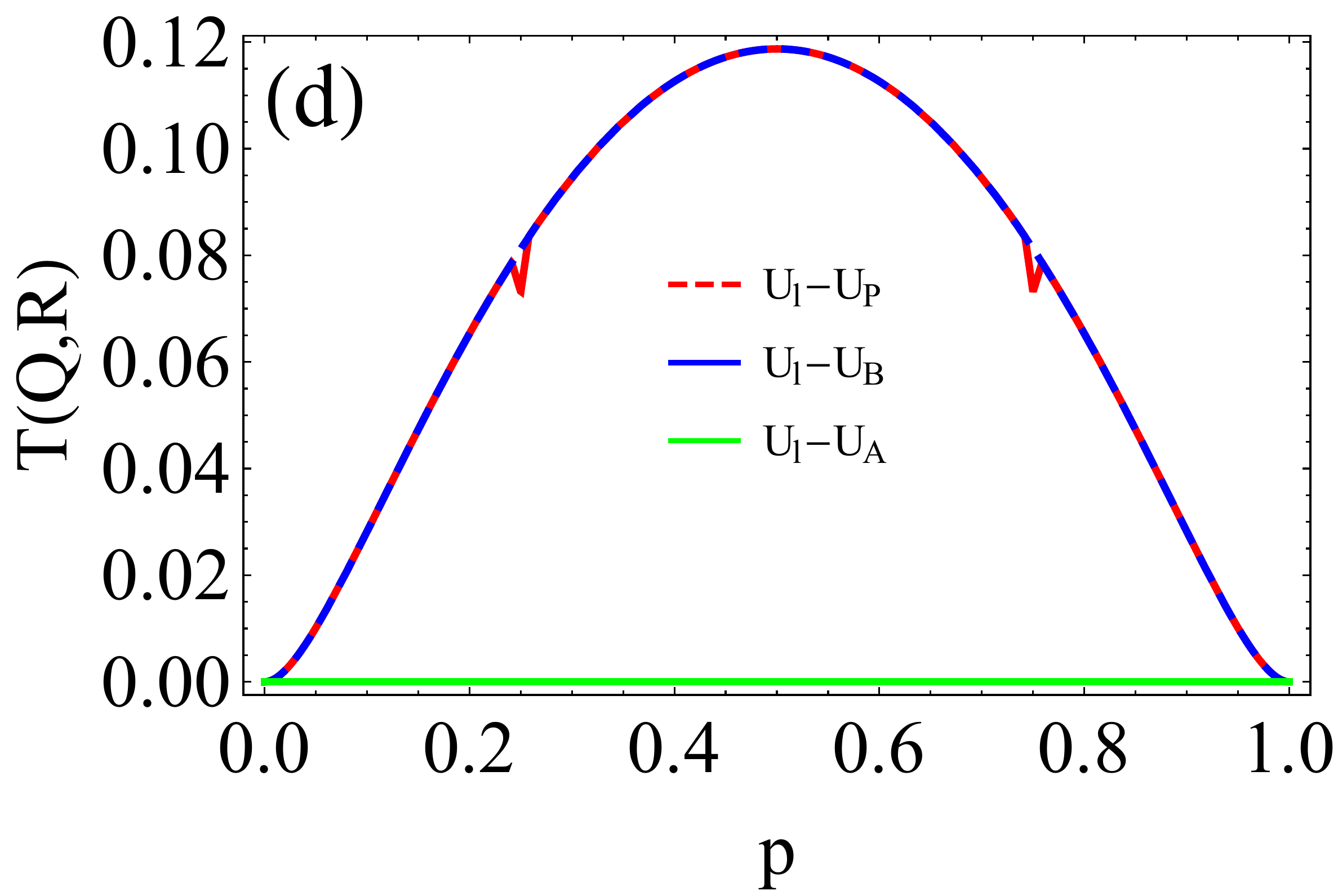}}
\caption{
	Dynamics of entropic uncertainty under BPF  channel, several kinds of lower bound and quantum correlation of the system  as a function of time parameter $p$   when the initial state is prepared with $ C_1=-0.5,C_2=0.4,C_3=0.8 .$ Graph (a): $U_l$ denotes the entropic uncertainty expressed in Eq. (\ref{E9}) , the  black dashed line denotes the bound proposed by Berta et al., the pink line denotes the Pati et al.'s bound, the yellow dashed line represents the bound by Adabi et al. The  graph (b) and(d): $U_A$ denotes the Adabi et al.'s lower bound; $U_B$ denotes the Berta et al.'s bound, $U_P$ denotes the Pati et al.'s bound. Graph (c):QD denotes quantum discord expressed in  Eq. (\ref{E9}) , $S^{A\vert B}_{\rm min}$ denotes the $A$'s minimal condition entropy in Eq. (\ref{E26}).}
		\label{fig:2}
\end{figure*}

	The four eigenvalues of matrix are
		\begin{align}
\lambda^{\rm BPF}_1=\frac{1}{4}(1 + C_1 - C_2 + C_3 - 2 C_1 p - 2 C_3 p), \nonumber\\
\lambda^{\rm BPF}_2=\frac{1}{4}(1 - C_1 + C_2 + C_3 + 2 C_1 p - 2 C_3 p),\nonumber\\
\lambda^{\rm BPF}_3=\frac{1}{4}(1 + C_1 + C_2 - C_3 - 2 C_1 p + 2 C_3 p),\nonumber\\
 \lambda^{\rm BPF}_4=\frac{1}{4}(1 - C_1 - C_2 - C_3 + 2 C_1 p + 2 C_3 p).	\label{}
	\end{align}

	Thus, the left-hand side of Eq. (\ref{E9}) can be given by
	
		\[U^{BPF}=-\frac{1}{2}\{-2\nu_1\rm arctanh(\nu_1)-\log_2(1+\nu_1)\]
			\[-\log_2(1-\nu_1)-(1+\nu_2)\log_2(1+\nu_2)+(\nu_2-1)\log_2(1-\nu_2),\]
		 where $\nu_1=C_1 - 2 C_1 p,\nu_2=C_3 - 2 C_3 p.$

	Similarly, the right-hand side (i.e., bound) of equation takes the form
	
	\begin{equation}
U^{\rm BPF}_b=-\frac{1}{4}\sum_{i=1}^{4}\zeta_i \log_2(\frac{\zeta_i}{4})
\label{E31}
\end{equation}
	
	where	$ \zeta_1=1 + C_1 + C_2 - C_3 - 2 C_1 p + 2 C_3 p, \zeta_2=(1 + C_1 - C_2 + C_3 - 2 (C_1 + C_3) p), \zeta_3=1 - C_1 - C_2 - C_3 + 2 (C_1 + C_3) p, \zeta_4=1 + C_2 + C_3 - 2 C_3 p + C_1 (-1 + 2 p)),$ we plot the diagrams of evolution of entropy uncertain relation and the existed	bounds in Figure \ref{fig:2} . Likewise,the dynamics of the entropic uncertainty expose   non-monotonic feature and  gradually close  , Berta et al.'s  bound   is largest, relatively that Pati et al.'s bound is between the other two, Adabi et al.'s bound is still perfectly coincident with the left-hand side of the uncertainty.And it is worth noting that different from AD channel , but in Figure \ref{fig:2}b  Pati's bound is  coincident with the right-hand side of uncertainty except channel parameter $p$ at 0.25 and 0.75.

 	\section{  A strategy to steer the uncertainty of measurement via a type of flipping operations and quantum weak measurement }
 	\label{S3}
In open systems, the effects of decoherence or dissipation are very significant?in fact, any quantum system is actually open and inevitably affected by its surrounding environment noises. We can deduce in this respect that when the environment and the system are coupled, the uncertainty will increase more or less. Thus, here naturally comes a question:how to manipulate the magnitude of the uncertainty with regard to a given system in order to reduce the impact of the external environment? Up to now, some authors presented some efficient methods to protect quantum states and quantum correlations from the noises,for example, filtering operator (FO) and  quantum weak
	measurement (QWM). We are curious about that whether such operations are valid to steer
 	the quantity of the uncertainty. After our investigation, we
	find the answer is positive. In the followings, we illustrate
	how these two methods achieve the reduction of the magnitude of the entropic uncertainty.
	
Firstly we  exploit a filtering operator on system experiences any noises to decrease the uncertainty , and the Kraus operator of FO can be written as  \cite{27,28}
	 \begin{equation}
	\hat{M}= \left(
	\begin{array}{cc}
	\sqrt{1-k} & 0  \\
	0  & \sqrt{k}\\
	\end{array}
	\right),
	\label{E32}
	\end{equation}
	where $\hat{M}$ represents the operator acting on particles $A$ , and $k$ is the measurement strength with $1>k>0$, and it can recover and increase entanglement to some extent \cite{29,30}. Practically,we will expound the de the details of how FO reduce the uncertainty of the measurement with pertain to two incompatibility in the follows .
	
	 After  the FO operation on the qubit $A$ to be detected, the  final state of the bipartite-system consisting AB can express as
	  \begin{equation}
	\rho^{M}_{AD}(t)=\frac{(\hat{M}\otimes I)\rho_{AB}(\hat{M}\otimes I)^\dagger}{[Tr(\hat{M}\otimes I)\rho_{AB}(\hat{M}\otimes I)^\dagger]}.
	\label{E33}
	 \end{equation}
	
	 After normalization, one can obtain the density matrix $  	\rho^{M}_{AD}(t) $ of the system
	 	\begin{align}
	\rho^{M}_{AD}(t)_{11}=&\frac{(1 + C_3 + d - C_3 d) (1 - k)}{2 (1 + d - 2 d k)},\nonumber\\
	\rho^{M}_{AD}(t)_{22}=&-\frac{(1 + C_3 (-1 + d) + d) (-1 + k)}{2 (1 + d - 2 d k)},\nonumber\\
	\rho^{M}_{AD}(t)_{22}=&-\frac{(1 + C_3 (-1 + d) + d) (-1 + k)}{2 (1 + d - 2 d k)},\nonumber\\
	\rho^{M}_{AD}(t)_{33}=&\frac{(-1 + C_3) (-1 + d) k}{2 (1 + d - 2 d k)},\nonumber\\
		\rho^{M}_{AD}(t)_{14}=&\rho^{M}_{AD}(t)_{14}=\frac{(C_1 - C_2)\sqrt{1-d}\sqrt{(1-k)k}}{2 (1 + d - 2 d k)},\nonumber\\
		\rho^{M}_{AD}(t)_{23}=&\rho^{M}_{AD}(t)_{32}=\frac{(C_1 + C_2)\sqrt{1-d}\sqrt{(1-k)k}}{2 (1 + d - 2 d k)}.
		 \end{align}
	 
 	 	   By calculating their eigenvalues, we can attain the corresponding von Neumann entropies as
	  			
	  				\begin{align}
	  		 H(\rho^{M}_{XB})=-\sum_{i=1}^{4}\lambda^{M}_{XBi}\log_2(\lambda^{M}_{XBi}),\nonumber\\
	  		 	  H(\rho^{M}_{ZB})=-\sum_{i=1}^{4}\lambda^{M}_{ZBi}\log_2(\lambda^{M}_{ZBi}),
	  			\end{align}	  	
	  				 	  with
	  				  	\begin{align}
	  				\lambda^{M}_{XB1}=  \lambda^{M}_{XB3}= \frac{1}{4}-\delta,\nonumber\\
	  				  \lambda^{M}_{XB2}=  \lambda^{M}_{XB4}=\frac{1}{4}+\delta,
	  				  \end{align}	
	  				  $\delta=\frac{\sqrt{(-1 + d) (C_3^2 (-1 + d) (1 - 2 k)^2 + 4 C_1^2 (-1 + k) k)}}{-4 + d (-4 + 8 k)}$ and $ \lambda^{M}_{ZBi}=	\rho^{M}_{AD}(t)_{ii}. $
	  				
	  				  In addition, we have
	  				
	  				  	\begin{equation}
	  				  H(\rho^{M}_{B})=-\sum_{i=1}^{2}\lambda^{M}_{Bi}\log_2(\lambda^{M}_{Bi}),
	  				  \label{E36}
	  				  \end{equation}
	  				    where 
	  				    	\begin{align}
	  				    \lambda^{M}_{B1}=&\frac{\sqrt{1 + d - 2 d k + C_3 (-1 + d) (-1 + 2 k)}}{2 + d (2 - 4 k)},\nonumber\\
	  				    \lambda^{M}_{B2}=&\frac{\sqrt{-1 + C_3 (-1 + d) (-1 + 2 k) + d (-1 + 2 k)}}{-2 + d (-2 + 4 k)}.
	  				    \end{align}
	  				    
	  				   As a  result, we can get the uncertainty of von Neumann entropy uncertainty is
	  				  	
	  				  \begin{align}
 	  				  U^M=&-\sum_{i=1}^{4}\lambda^{M}_{ZBi}\log_2(\lambda^{M}_{ZBi})-2\sum_{i=1}^{2}\lambda^{M}_{XBi}\log_2(\lambda^{M}_{XBi}) \nonumber \\
	  				  &+2\sum_{i=1}^{2}\lambda^{M}_{Bi}\log_2(\lambda^{M}_{Bi})
	  				  \label{E37}
	  				  \end{align}

\begin{figure*}
\centering
{\includegraphics[width=7cm]{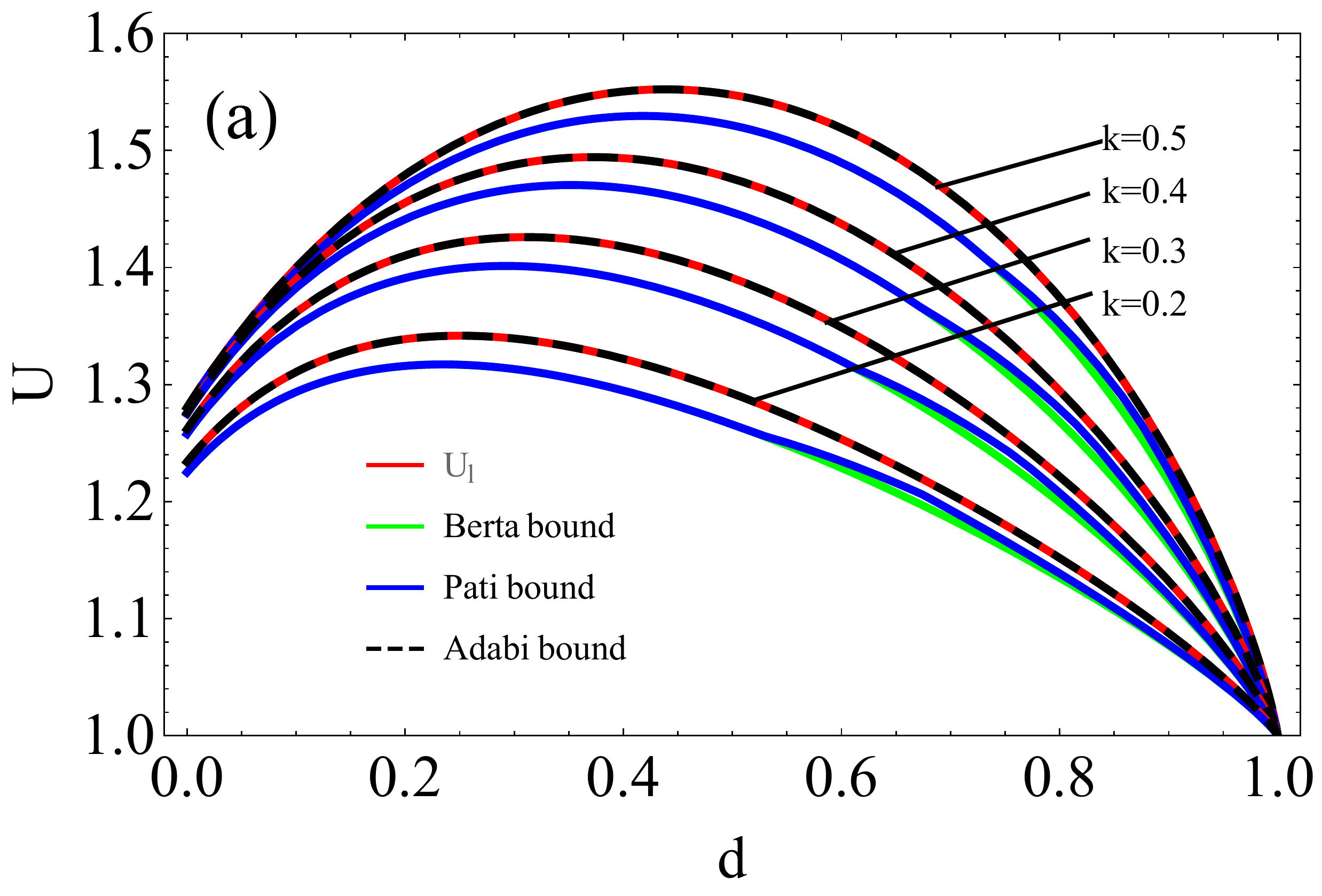}} \ \ \ \
{\includegraphics[width=7cm]{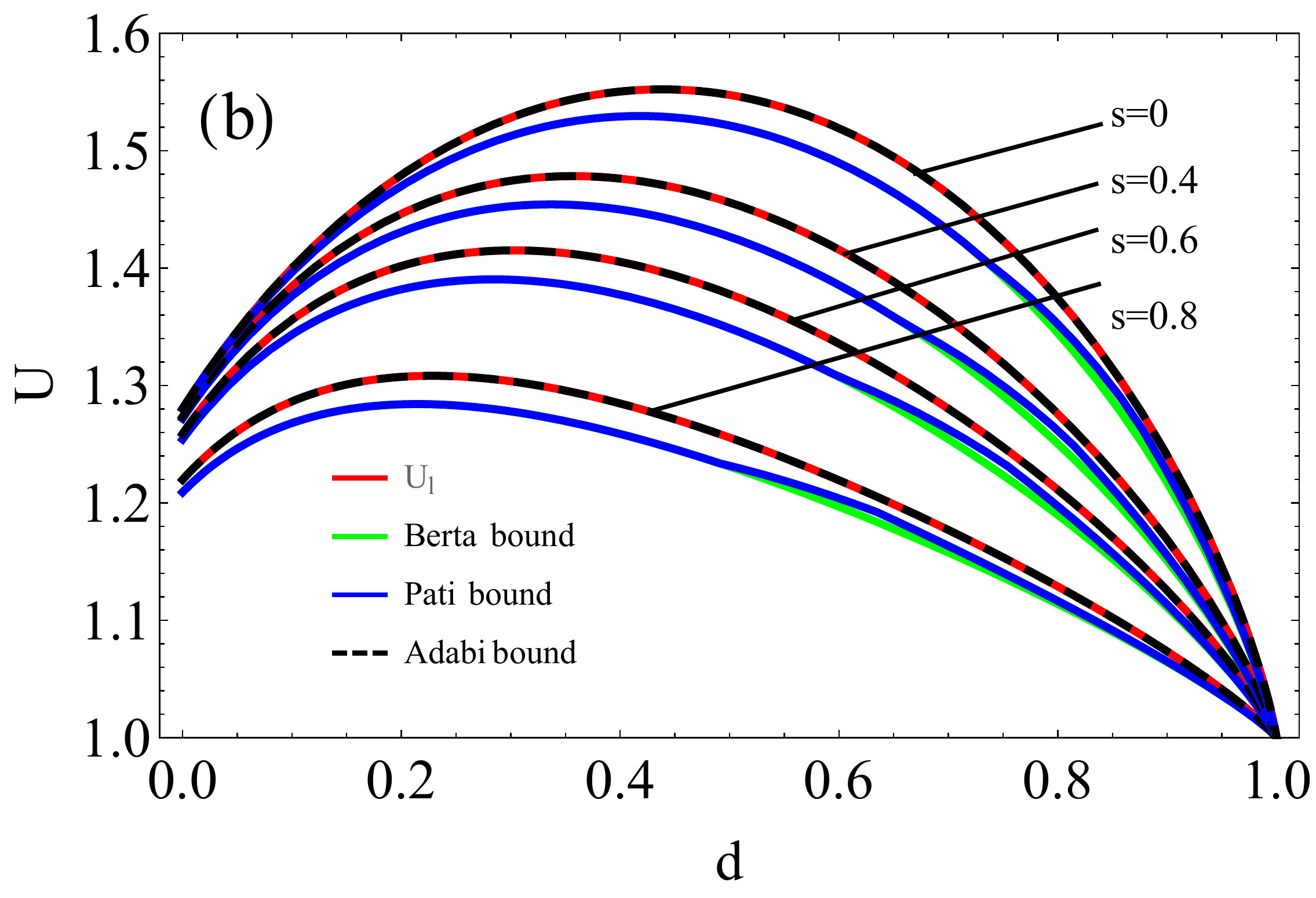}}
\caption{The entropy uncertainty $U$ is plotted as a function of the operation strength $k,s$ and the states parameter $d$. the initial state under AD channel is prepared same with $ C_1=-0.5,C_2=0.4,C_3=0.8 .$ In the Gaph (a): the entropic uncertainty with $d$ with different via filtering operator strengths with an initial Bell state constructed with the $k_1 =0.2,k_2=0.3,k_3=0.4,k_4=0.5,$ Graph (b): the entropic uncertainty with $d$ with an initial Bell-diagonal state  constructed with the $s_1 =0,s_2=0.4,s_3=0.6,s_4=0.8$ under   different weak
	 		measurement strengths.}
	 		\label{fig:3}
\end{figure*}

	 In our consideration, the qubitt $A$ be measured suffers from a  environment noise, while qubit $B$ as quantum memory is free from any noises. 	Moreover, We display in Fig. \ref{fig:3} how the measured entropy-based uncertainty  changes with the state parameter $d$ .
	
%
%

	Furthermore, we can exploit the quantum weak measurement to reduce the amount of the measuring uncertainty, The quantum weak measurement is a non-trace-preser
	  map (NTPM) operation, which Kraus operator can be expressed by \cite{31,32,33}
	 	 \begin{equation}
	 \hat{W }= \left(
	 \begin{array}{cc}
	1 & 0  \\
	 0  & \sqrt{1-s}\\
	 \end{array}
	 \right),
	 \label{E38}
	 \end{equation}
	 where $s$ indicates the strength of the weak measurement
	 operation and satisfies $0 <s <1$. When this operation is
	 acted on qubit A, the final state of the Quantum system would be written by
	 \begin{equation}
	 \rho^{W}_{AD}(t)=\frac{(\hat{W}\otimes I)\rho_{AB}(\hat{W}\otimes I)^\dagger}{[Tr(\hat{W}\otimes I)\rho_{AB}(\hat{W}\otimes I)^\dagger]}
	 \label{E39}
	 \end{equation}
	
	 As same , we consider this case where $A$ experiencing AD
	 channels. After the operation acts on $A$ to be observed, the
	 post-operation density matrix of becomes
	 	\begin{align}
	\rho^{W}_{AD}(t)_{11}=&-\frac{1 + C_3 + d - C_3 d}{4 (1 + 1/2 (-1 + d) s)},\nonumber\\
	\rho^{W}_{AD}(t)_{22}=&-\frac{1 + C_3 (-1 + d) + d) (-1 + s)}{4 (1 + 1/2 (-1 + d) s)},\nonumber\\
		\rho^{W}_{AD}(t)_{33}=&-\frac{(-1 + C_3) (-1 + d) (-1 + s)}{4 (1 + 1/2 (-1 + d) s)},\nonumber\\
			\rho^{W}_{AD}(t)_{44}=&\frac{(1 + C_3) (-1 + d) (-1 + s)}{4 (1 + 1/2 (-1 + d) s)},\nonumber\\
				\rho^{W}_{AD}(t)_{14}=&\rho^{W}_{AD}(t)_{14}=\frac{(C_1 - C_2)\sqrt{1-d}\sqrt{1-s}}{4 (1 + 1/2 (-1 + d) s)},\nonumber\\
						\rho^{W}_{AD}(t)_{23}=&\rho^{W}_{AD}(t)_{32}=\frac{(C_1 + C_2)\sqrt{1-d}\sqrt{1-s}}{4 (1 + 1/2 (-1 + d) s)}.\nonumber\\
	 \end{align}

 	and we can obtain the von Neumann entropy of state can be calculated as
 
 	\begin{align}
  H(\rho^{W}_{XB})=&-\frac{1}{2}[\frac{1}{2}\sum_{i=1}^{4}\log_2(\lambda^{T}_{XB_i})\nonumber\\
  &+4(\lambda^{T}_{XB_2}-1)\rm arccoth\frac{1}{4(\lambda^{T}_{XB_2})} -\eta \rm arctanh(\eta) ],\nonumber\\
   H(\rho^{W}_{ZB})=&-\sum_{i=1}^{4}\lambda^{T}_{ZBi}\log_2(\lambda^{T}_{ZBi}),
   \label{E40}
 \end{align}
  	with
	 	 	\begin{align}
	 \lambda^{W}_{XB1}=\lambda^{W}_{XB3}= \frac{1}{4}-\epsilon,\nonumber\\
	 \lambda^{W}_{XB2}=\lambda^{W}_{XB4}= \frac{1}{4}+\epsilon,
	 	\end{align}
	 	 $\epsilon=\frac{\sqrt{(-1 + d) (4 C_1^2 (-1 + s) + C_3^2 (-1 + d) s^2)}}{8 + 4 (-1 + d) s}	,\eta=\frac{2C_3 (-1 + d) s }{2 + (-1 + d) s}$ and $ \lambda^{W}_{ZBi}=	\rho^{W}_{AD}(t)_{ii}. $

	  From the Fig. \ref{fig:3}, one can easily get that the measured uncertainty decrease  monotonically  with  the increase of the weak measurement strength $ k$ and the FO operation intensity $ s$
, which reflects that such an optimal measurement choice  to reduce the measured uncertainty is very effective, which is pretty necessary during practical quantum-information-processing. Put another way, the entropy uncertainty can be manipulated by  optimal strategy satisfactorily, best combination of a FO and QWM .
%

\begin{figure*}
\centering
{\includegraphics[width=7cm]{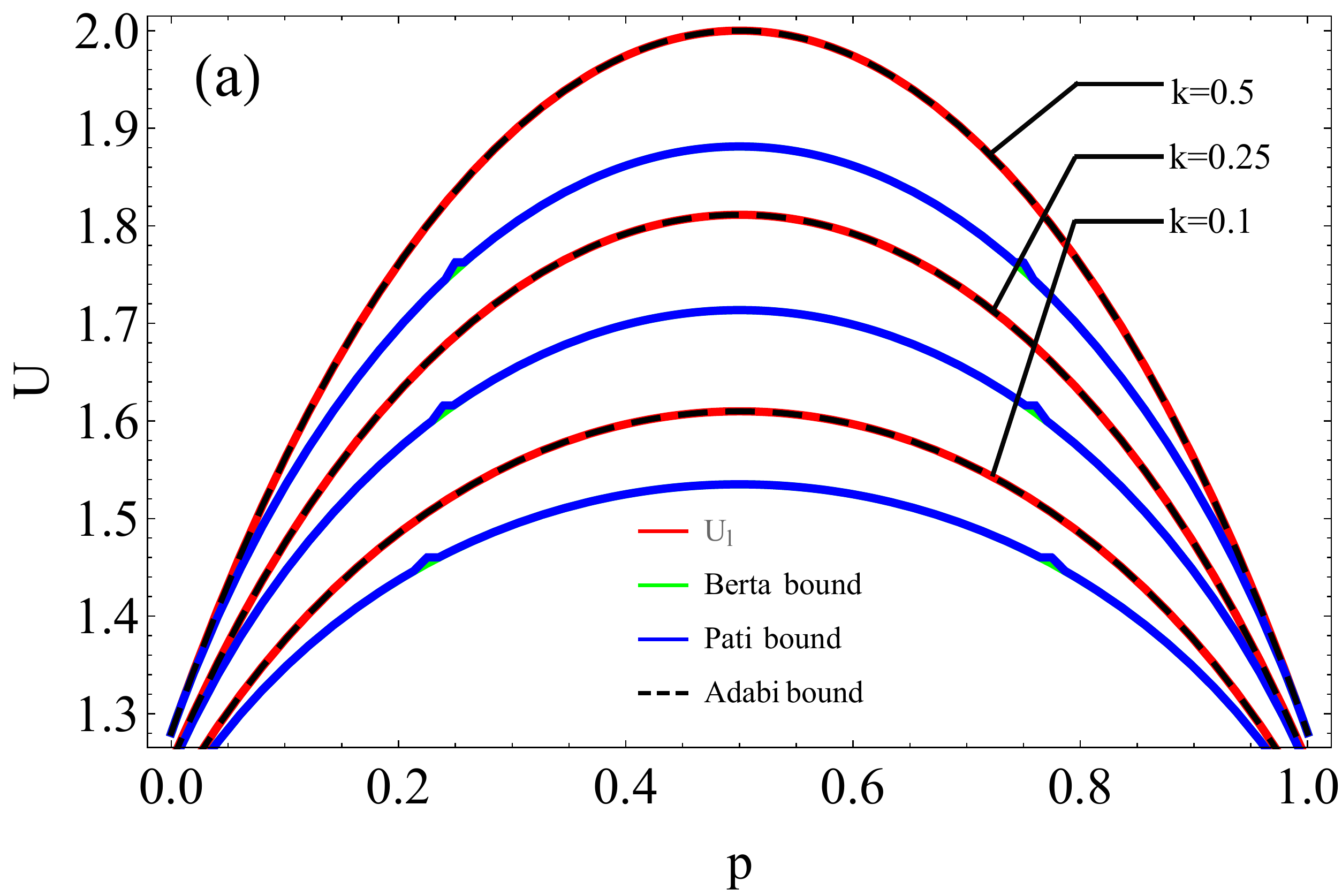}} \ \ \ \
{\includegraphics[width=7cm]{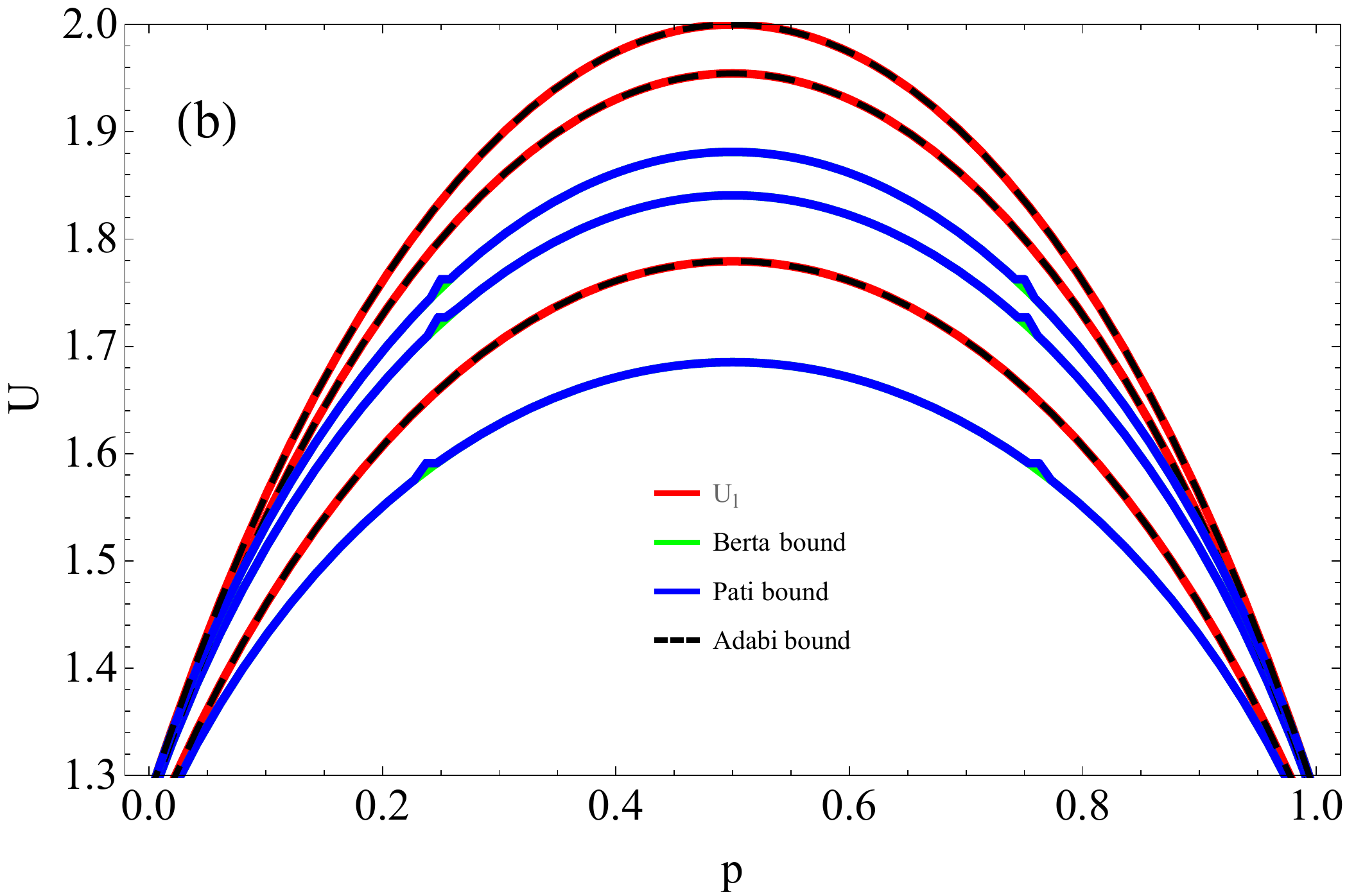}}
\caption{
	The entropy uncertainty $U$ is plotted as a function of the operation strength $k,s$ and the states under BPF noise with parameter $p$. Gaph (a): the entropic uncertainty with different via filtering operator strengths with an initial Bell state constructed with the $k_1 =0.1,k_2=0.25,k_3=0.5,$ Graph (b): the entropic uncertainty with $d$ with an initial Bell-diagonal state  constructed with the $s_1 =0.7,s_2=0.4,s_3=0$ under   different weak	measurement strengths.}
\label{fig:4}
\end{figure*}

	  In addition, from the Fig. \ref{fig:4} ,  obviously,this is easy to achieve that the state parameter $p$ and $s$ will also monotonically reduce  the uncertainty of BPF channel.
	 This fact is because because we adjust the state parameters to cause higher purity of the system, so the higher the purity, the smaller the uncertainty. 
	 This shows that in order to obtain lower measurement uncertainty, we can carry out this scheme in a composite system, 
	therefore, we can manipulate this operation to still gain high-precision measurements when encountering various realistic noises caused by the external environment.

	\section{  Application}
	\label{S4}
		\subsection{Entanglement witness }
	
		 One of the significant form of quantum correlation is quantum entanglement, And has produced very important applications in both communication and quantum information  , such as quantum dense coding \cite{34}, quantum telecloning \cite{35}, teleportation \cite{36},  quantum computing \cite{37}, quantum electrodynamics (QED) \cite{38,39,40,41,42,43} ,and so on. Therefore, in the field of quantum information processing, how to determine whether quantum states are entangled has become a very important topic. Considering this , we briefly introduce an effective criterion based on uncertainty relations in the current scenario.
		Technically, $S(A\vert B) \leq  0$ shows the bound can be reduced contrast to the previous one in Eq. (\ref{E8}),  Therefore it is often used as  an indicator of entanglement witness.
	
\begin{figure*}
\centering
{\includegraphics[width=7cm]{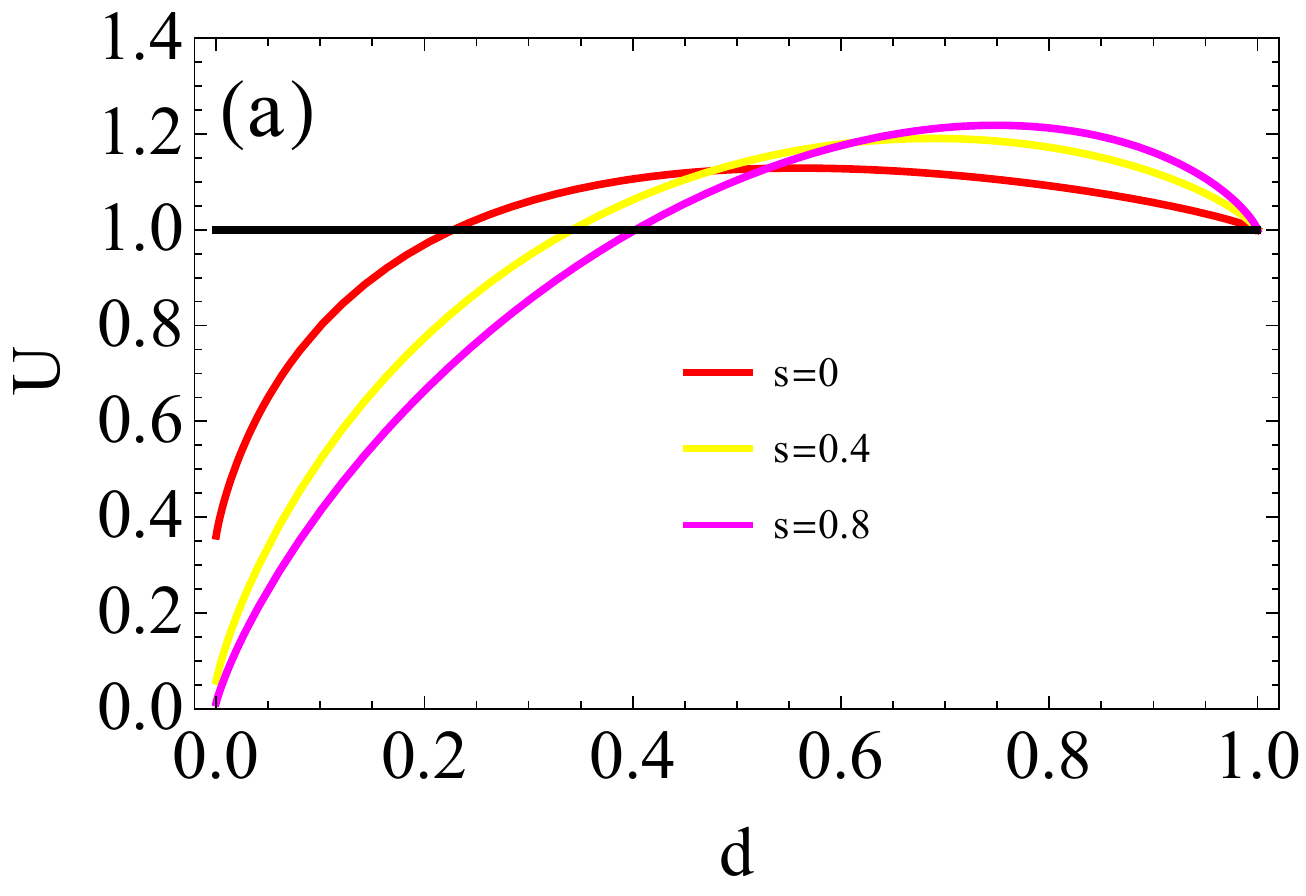}} \ \ \ \
{\includegraphics[width=7cm]{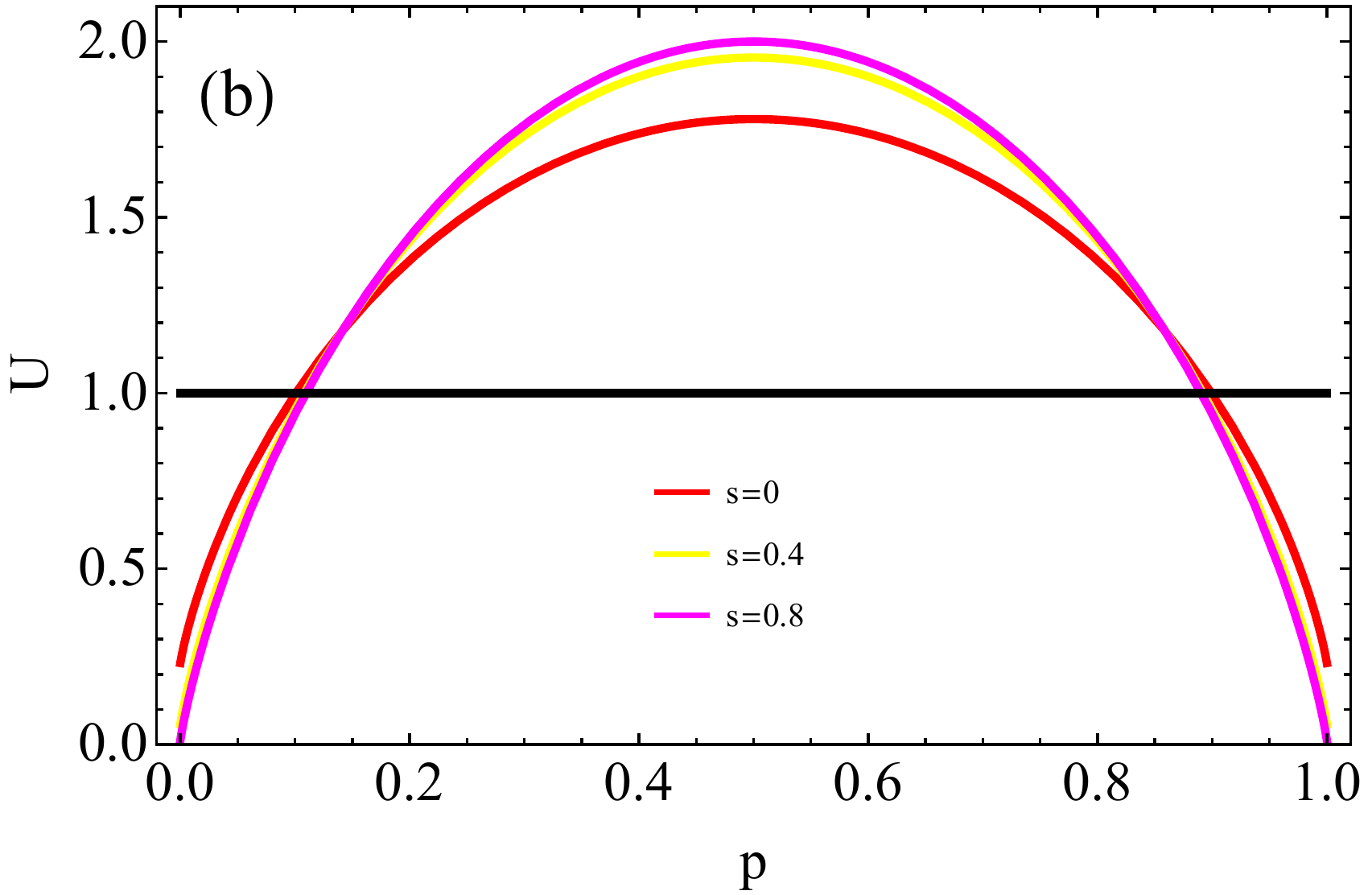}}
\caption{Evolution of entropic uncertainty $U$ when the particle $A$ is under AD and BPF channel. Graph (a):  Evolution of $U$ with respect $d$ to  when $s$ takes different values. Graph (b): Evolution of $U$ with respect $p$ to  when $s$ takes different values, the pink line is plotted with s = 0, the yellow line is plotted
			with s = 0.4, the red line is plotted with s = 0.8 and the red line with p = 0.5. Here we set $C_1 =-1, C_2 =1, C_3 = 1$.}
				\label{fig:5}
\end{figure*}

	According to the Eq. (\ref{E9}), if the inequality satisfy $U=S(\hat{P}\vert B)+S(\hat{R}\vert B)\leq \log_2\frac{1}{c}=1$ which means  $A$ entangle with $B$. In Fig. \ref{fig:5},  we shown the relationship between $S(\hat{P}\vert B)+S(\hat{R}\vert B)$ and $\log_2\frac{1}{c}=1$ with the maximal purity of initial  , therefore,by a rigorous calculation one can obtain the entanglement condition via $S(\hat{P}\vert B)+S(\hat{R}\vert B)<1$ 
	
%
%

Here the case when particle A under the AD and BPF channel . By setting $C_1 =-1, C_2 =1, C_3 = 1$, the initial state has maximal purity, and consider the entropic uncertainty under quantum weak measurement Eq. (\ref{E38}), we then can get the
relationship with $d, p$ and $s$ ,when  noisy streng of the AD channel
	
	\begin{equation}
d<d_m,
\end{equation}
	 for instance, we can calculate the $d_m$  and its value is
	 $d_m = 0.4058$ with $s = 0,$ and BPF noisy strength
	 	\begin{equation}
[0,p_m)\cup(1-p_m,1],
	 \end{equation}
	when $U<1$ , the $p_m$  is
	$p_m = 0.1125$ with $s = 0,$ in an environment where the initial system reaches maximum purity.
	
		\subsection{Quantum channel capacities }
		Quantum channels generally describe the situation in which any physical process acts on a quantum system \cite{44}, including  achieving non-eavesdropping secure communication and sending secret classical information . And quantum channel capacity means the maximal amount of information transfer via quantum channel, which can be written as follows \cite{45}
			\begin{equation}
		C(\rho_{AB})=\rm max\{I(A:B)\},
		\label{E49}
		\end{equation}

	where $ I(A:B)$  indicate the mutual information of system $\rho_{AB}$. Here, we are able to build the connection between the uncertainty's bound and the channel capacity  , which can be derived as
		\begin{equation}
U_b=\log_2\frac{1}{c}+S(A\vert B)=1+S(A\vert B) ,
\label{E46}
	\end{equation}
	by substituting Eq. (\ref{E49}) and Eq. (\ref{E14}) into the above equation, 
	we can obtain the channel capacity $C(\rho_{AB})$ can be written as
	  	\begin{equation}
	C(\rho_{AB}) = \rm max[S (\rho_{A})-U_b +1]  ,
	\label{E47}
	  \end{equation}
	  which shows the connection between the uncertainty relation and the quantum channel capacity. So we can get the exact expression of capacity as that
 	\begin{equation}
C(\rho_{AB})=\sum_{i=1}^{4}\lambda^{AB}_i\log_2\lambda^{AB}_i-\sum_{i=1}^{2}\lambda^{A}_i\log_2\lambda^{A}_i-\sum_{i=1}^{2}\lambda^{B}_i\log_2\lambda^{B}_i  ,
\label{E48}
 \end{equation}
 and the $\lambda_i^{AB}$ are the eigenstates of state $\rho ^{AB}$, which have been given above. Clearly, we get the eigenstates of bipartite state under AD channel $\rho_{A}$and $\rho_{B}$ are $\lambda^A_{AD_1}=\frac{1-d}{2},\lambda^A_{AD_2}=\frac{1+d}{2},\lambda^B_{AD_1}=\lambda^B_{AD_2}=\frac{1}{2},$ and under BPF channel $\lambda^A_{BPF_1}=\lambda^A_{BPF_2}=\lambda^B_{BPF_1}=\lambda^B_{BPF_1}=\frac{1}{2},$ and we thus have
 
 	\begin{align}
 &C_{AD}(\rho_{AB})=\sum_{i=1}^{4}\lambda^{AB}_{AD_i}\log_2\lambda^{AB}_{AD_i}-\sum_{i=1}^{2}\lambda^{A}_{AD_2}\log_2\lambda^{A}_{AD_i}+1  ,\nonumber\\
 &C_{BPF}(\rho_{AB})=\sum_{i=1}^{4}\lambda^{AB}_{BPF_i}\log_2\lambda^{AB}_{BPF_i}+2 .
   \label{E49}
 \end{align}
 	
%
 \begin{figure*}
\centering
{\includegraphics[width=7cm]{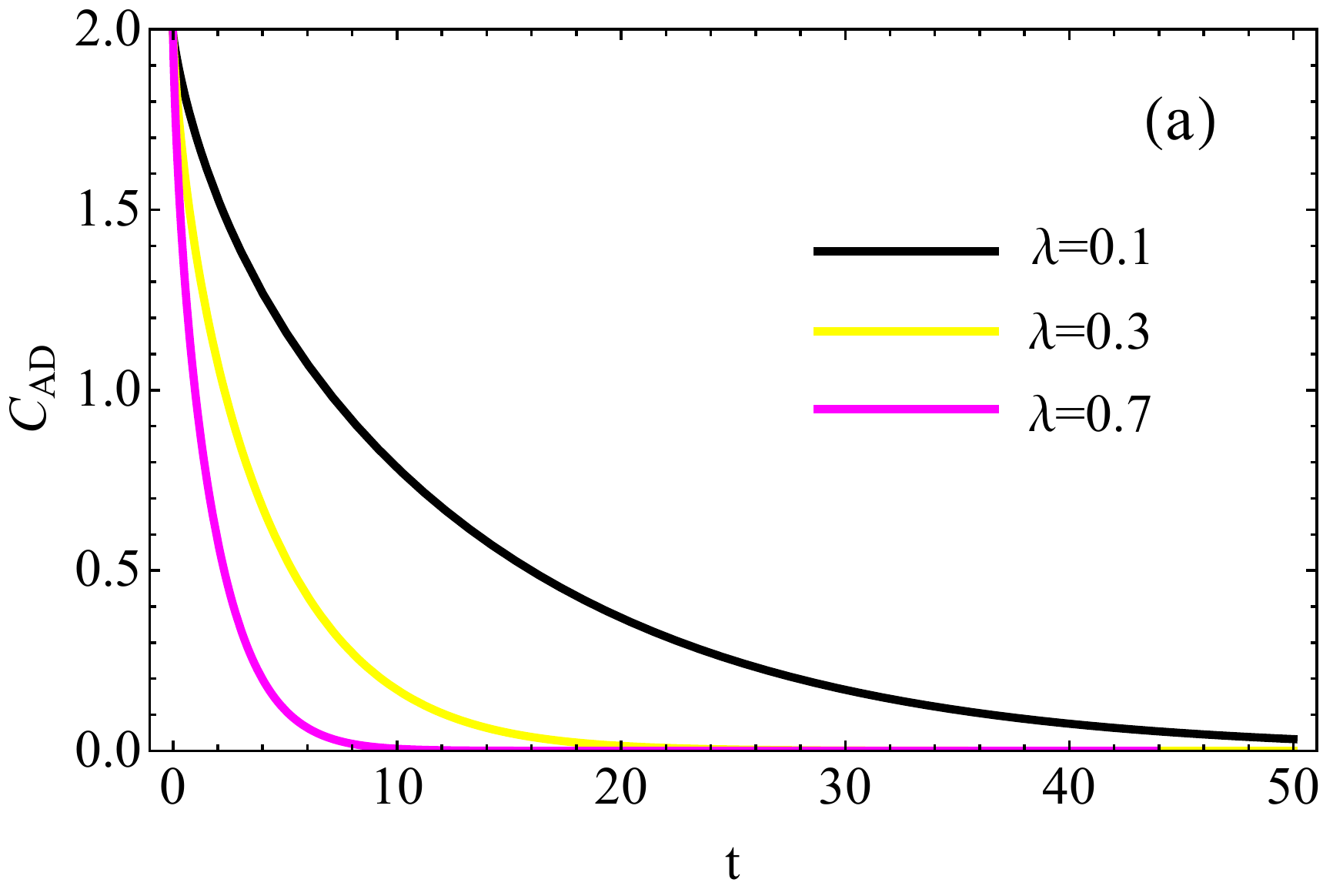}} \ \ \ \
{\includegraphics[width=7cm]{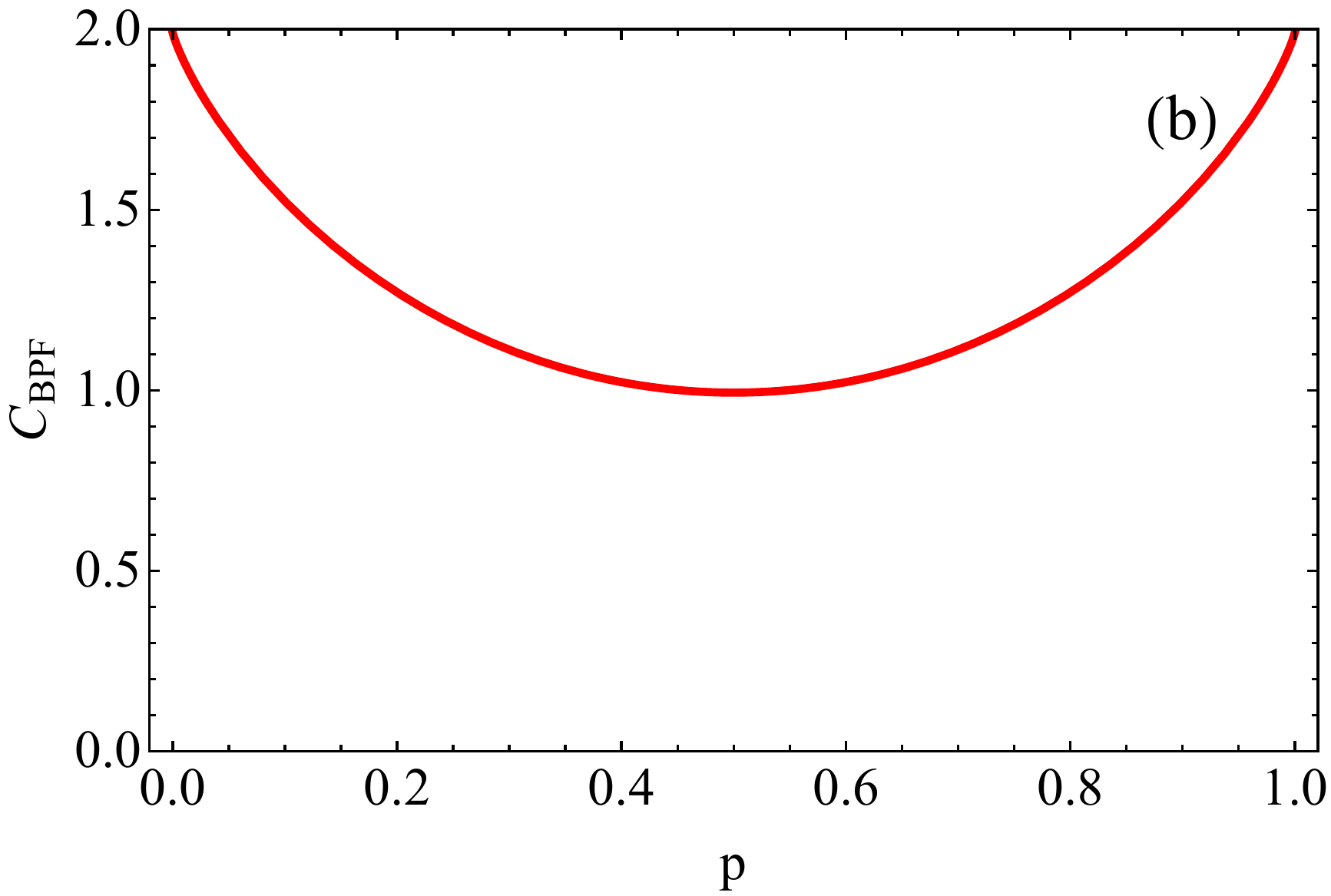}}
\caption{Evolution of channel capacity with maximal purity condition $(C_1=C_2=1,C_3=-1). $ Graph (a):The quantum channel capacity $C_{AD}$ a function of time $t$ and parameter $\lambda$ when takes different values, the black line is drawed with $\lambda=0.1$, the yellow line is drawed with $\lambda=0.3$ and the pink line is drawed with $\lambda=0.7. $ Graph (b):The quantum channel capacity $C_{BPF}$ a function of $p .$}
 	   	 	  	   	\label{fig:6}
\end{figure*}
	  	
   In order to show the relationship between the capacity and the bound more explicitly, the channel capacity we provide is a function of time $t,p$ and parameter $\lambda$ in terms of setting maximal purity condition $(C_1=C_2=1,C_3=-1), $  following the Fig. \ref{fig:6} (a), it can be clearly found that the evolution trend of capacity is similarly tocorrelated with the bound-reducing ,but in Fig. \ref{fig:6} (b) , the capacity reducing firstly, when the parameter equal 0.5, the capacity anti-correlated grow to maximum bound. Considering these, due to the intrinsic connection between channel capacity and the uncertainty's bound, the channel capacity can be obtained through analysis in the architecture.
	\section{  Conclusions}
	\label{S5}
 We investigate the dynamics of the quantum memory-assisted entropic uncertainty of bipartite  system to be measured are coupled with two different external noises in open environment. We considered two different noise situations that 	particle A and B entangle in  Bell-diagonal state suffers the AD or BPF channels.And we  choose mutual-noncommuting bases observation of Pauli operators $\sigma_x$ and $\sigma_z$ as the incompatibility to investigate dynamical properties of three different  tighten  bounds .Through numerical analysis, we can draw the following conclusions: (\uppercase\expandafter{\romannumeral1}) When particles A  suffer from the different noise of environments , uncertainty and bunds have non-monotonic feature and it can be obtained  that its uncertainty will drop to the lower bound for a long time.  (\uppercase\expandafter{\romannumeral2}) The evolution of different bounds are as expected that the bound of Adabi et al. is tighter than that of Pati et al. ,while the Pati et al.'s bound is tighter than Berta et al.'s bound. (\uppercase\expandafter{\romannumeral3})   the correlated between quantum discord of the systemand bund is not completely negtive. This can be explained as that the evolutionary characteristics of  bound is not only dependent of QD, but also  the A's minimal conditional entropy $S^{A\vert B}_{\min}$ , which has been derived forenamed. We infer that the quantum memory effect can essentially reduce the increase of uncertainty and its lower bound. What?s more, we have put forward two	effective operations to manipulate the magnitude of the measurement's uncertainty under the open system via deriving quantum weak measurement and filtering operation separately, 	it proves that the above operations can effectively suppress the inflation of the uncertainty, and also reduce the magnitude to a large extent  . At last,we consider the applications on entanglement witness and quantum channel capacity. Therefore, we claim	the  investigations we did  may be of great significance for the realization of practical quantum measurement experment, and  maymight be beneficial to help us understand the dynamical behavior and operation  of entropy uncertainty in open systems.

\renewcommand\refname{Reference}

\end{document}